\documentclass[fleqn,usenatbib]{mnras}

\usepackage[T1]{fontenc}

\usepackage{placeins}
\usepackage{afterpage}

\DeclareRobustCommand{\VAN}[3]{#2}
\let\VANthebibliography\thebibliography
\def\thebibliography{\DeclareRobustCommand{\VAN}[3]{##3}\VANthebibliography}

\usepackage{graphicx}	
\usepackage{amsmath}	
\usepackage{amssymb}	
\usepackage{gensymb}
\usepackage{xcolor}
\usepackage[]{multirow}
\usepackage{ulem}
\usepackage{subfigure}
\usepackage{comment}
\newcommand\sunmass{M_\odot}

\usepackage{newtxtext,newtxmath}

\usepackage[switch, modulo]{lineno}

\title[Modelling the variable emission states of $\gamma$-ray emitting NLS1 galaxies]{Modelling the variable emission states of $\gamma$-ray emitting Narrow-Line Seyfert 1 galaxies}

\author[A. Luashvili et al.]{
Anna Luashvili,$^{1}$\thanks{E-mail: anna.luashvili@obspm.fr}
Catherine Boisson,$^{1}$
Andreas Zech,$^{1}$
Maialen Arrieta-Lobo,$^{2}$
Daniel Kynoch$^{3}$
\\
$^{1}$Laboratoire Univers et Théories, Observatoire de Paris, Université PSL, Université Paris Cité, CNRS, F-92190 Meudon, France\\
$^{2}$Sorbonne Université, Centre National de la Recherche Scientifique UMR 8256, Paris, France\\
$^{3}$School of Physics and Astronomy, University of Southampton, University Road, Southampton, SO17 1BJ, UK
}

\date{Accepted XXX. Received YYY; in original form ZZZ}

\pubyear{2022}

\begin{document}
\label{firstpage} 
\pagerange{\pageref{firstpage}--\pageref{lastpage}}
\maketitle

\begin{abstract}
$\gamma$-ray-emitting narrow-line Seyfert 1 galaxies ($\gamma$-NLS1) constitute an intriguing small population of Active Galactic Nuclei with $\gamma$-ray emission resembling low power flat-spectrum radio quasars (FSRQ), but with differing physical properties. They are jetted, $\gamma$/radio-loud Seyfert galaxies, with relatively low black hole masses, accreting at exceptionally high, near-Eddington rates. Certain of these sources exhibit highly variable emission states on relatively short time scales, the physical origin of which remains elusive. In this work, varying emission states of two bona-fide NLS1s, 1H\,0323+342 and PMN\,J0948+0022, and one little studied FSRQ/intermediate object, B2\,0954+25A, are examined. For each source, we analyzed quasi-simultaneous multiwavelength data for different states of $\gamma$-ray activity and present the results of their broad-band emission modelling, taking into account all available physical constraints to limit the range of the model parameters. Two different scenarios are discussed, in the framework of a one-zone leptonic model, where the high energy emission is due to the inverse Compton scattering of the disc and broad line region (BLR) or torus photons by relativistic electrons within the jet. The transition from low to high state is well described by variations of the jet parameters, leaving the external photon fields unchanged. The parameterisation favours an emission scenario with particle injection on a stationary shock inside the jet. When considering all physical constraints, the disc \& BLR scenario is preferred for all three sources. We use the multi-epoch modelling to characterize total jet powers and discuss the intrinsic nature of $\gamma$-NLS1 galaxies and FSRQs.

\end{abstract}




\section{Introduction}

Since their identification as a particular class of broad-line active galactic nuclei (AGNs), narrow-line Seyfert 1 galaxies (NLS1s) have drawn substantial attention. NLS1s share some optical features with typical broad-line Seyfert 1s, but have narrower permitted emission lines (FWHM $(H\beta)< 2000 \, \mathrm{km} \,\mathrm{s}^{-1}$), pointing to a relatively low mass  black hole (BH) ($\sim 10^6-10^8\sunmass$), fairly intense FeII and OI emission, and weak [O III] lines with a ratio [O III] to H$\beta < 3$ (\citet{Osterbrock_Pogge_1985}, \citet{Goodrich_1989}). In the X-rays, NLS1s also have extreme properties such as a remarkably soft X-ray excess ($0.5-2$ keV) or rapid and large-amplitude observed variability  (\citet{Boller_1996}). These X-ray characteristics that set NLS1s apart from other Seyfert types are mainly due to their high Eddington accretion rates ($L / L_{Edd} \sim 0.1-10$, \citet{Boroson_1992}; \citet{Boller_1996}) and their low-mass black holes with regards to Seyfert 1s. NLS1s are mostly radio-quiet, like Seyfert 1s, although a small fraction, $\sim$ 4--7\%, have been found to be radio-loud with a flat radio spectrum (see e.g. \citet{Komossa_2006}, \citet{Rakshit_2017}).

The detection of a small fraction of the radio-loud NLS1 galaxies by the \textit{Fermi} Gamma-Ray Space Telescope (\citet{Abdo_2009_Discovery}) has led to a new $\gamma$-loud class of active galactic nuclei (AGNs) that joins the well-known $\gamma$-ray emitting radio galaxies and blazars.
Due to their low black hole masses compared to blazars, NLS1s were not expected to harbor relativistic jets that could create such high-energy emission. The fact that this new type of $\gamma$-loud AGN features both blazar- and Seyfert-like characteristics could bring us closer to a better understanding of AGN unification, beyond the effect of their jet orientation, to describe different AGN types and provide new insight into the physics of the formation and evolution of radio jets under high accretion-rate conditions.
About twenty objects of this class have now been reported in the high energy (HE) band with {\it Fermi}-LAT, eight of them are listed in the Fourth Fermi-LAT
source catalog (4FGL; Abdollahi et al. 2020) (see \cite{Paliya_2019}, \cite{Foschini_2020_jetted_NLSy1} for a recent review).  They present both similarities and differences with the more powerful jetted AGNs, notably flat-spectrum radio quasar (FSRQs), see e.g. \citet{Foschini_2020_jetted_NLSy1} and \citet{D'Ammando_2019}.

Studies investigating the intrinsic power of relativistic jets in $\gamma$-NLS1s, and comparing them with the blazar class (e.g. \cite{Foschini_2015}, \cite{Paliya_2019}), are not easily comparable to each other, since they are based on different assumptions and methods, making it difficult to draw clear conclusions. 

The evidence of $\gamma$-ray emission poses the question of the location of the $\gamma$-ray emitting zone and the contribution of the $\gamma$-$\gamma$ absorption within the broad line region (BLR), since such objects are detected by {\it Fermi}-LAT but not by existing imaging Cherenkov telescopes, which have typical energy thresholds of at least a few 10\,GeV. Some of the $\gamma$-NLS1s show strongly changing spectral properties, in some cases even switching between Seyfert-type (accretion disc dominated) to blazar type (completely jet-dominated) states (e.g. \cite{D'Ammando_2015}). A better understanding of the mechanism behind varying emission states might be a key to constrain the overall emission process and the location of the emission region.
Hence, detailed studies involving multi-epoch spectral energy distribution (SED) modelling are required for better understanding of the nature of $\gamma$-NLS1 galaxies. 

The interest $\gamma$-NLS1 galaxies gained thanks to their intriguing physical properties led to intensive multi-wavelength (MWL) monitoring of these sources. Soon after the first detection in $\gamma$-rays, PMN J0948+0022 showed a spectacular HE flare in july 2009 (\cite{Foschini_2011}), whose SED modelling strengthened the similarities of these sources with powerful FSRQs.

Among the variable $\gamma$-ray detected NLS1s known to date, we have chosen two extensively studied sources, for which very well sampled MWL data sets exist: the nearest one being 1H\,0323+342, at a redshift of z = 0.0625 (\cite{Landt_2017}), and the highly variable PMN\,J0948+0022 at a higher redshift (z = 0.5846 \cite{Zhou_2003}). This selection is complemented by the source B2\,0954+25A, an FSRQ with similar SED and redshift as PMN\,J0948+0022 (see \citet{D'Ammando_2015}). We have modelled the SED of different activity states (low/average and flaring) with a one-zone model including synchrotron and self-Compton emission, as well as external Compton emission on the different photon fields inside the source. Two scenarios have been tested: one where the high-energy range is dominated by up-scattering of photons from the broad-line region and another one where the up-scattering of photons from the dusty torus is the most significant. 

Exploiting the available observational constraints concerning the source dimensions, energetics and variability time scales, we propose a common physical characterization of the variability and the jet power of these sources. 

Section \ref{section_model} provides the description of the adopted multi-component model and section \ref{sect_physical_constraints} lists the physical constraints we apply to reduce the available parameter space. Sections \ref{sect_1H}, \ref{sect_PMN} and \ref{sect_B2} present the individual source characteristics, the data sets and the results of our broad-band SED modelling for the three considered objects, 1H\,0323+342,  PMN\, J0948+0022 and B2\,0954+25A, respectively. The results of the modelling are discussed in Section \ref{discussion_section}, where we attempt a physical characterization of the mechanisms behind the varying broad-band emission and the source energetics of the two $\gamma$-NLS1s and compare them to the FSRQ under study.

Throughout the paper we adopt a $\Lambda$-CDM cosmology with $ \rm{H_0=70\ km s^{-1} Mpc^{-1}}$, $\Omega_m=0.3$, $\Omega_\Lambda=0.7$ \citep{Spergel2003}.

\section{Description of the code and the general approach} \label{section_model}

While a synchrotron self-Compton (SSC) model quite accurately describes the SEDs of most BL Lac type objects, for FSRQs, which exhibit a strong Compton dominance, $\gamma$-ray emission is thought to be produced by the external Compton (EC) mechanism on an ambient field of soft seed photons external to the jet. This also seems to be the case for $\gamma$-NLSy1 galaxies.
Seed photons relevant for one-zone models are the direct radiation from the accretion disc and the up-scattered radiation in the X-ray corona, the reprocessed radiation from BLR clouds, or from the dusty torus (e.g. \cite{Dermer_Schlickeiser_Mastichiadis_1992, Ghisellini_Tavecchio_2009}.

The multi-component code we have developed is based on the model described by \citet{Cerruti_2013} and \citet{Dermer_2014}. In the present version, the dust torus, disc and corona components are described following \citet{Ghisellini_Tavecchio_2009}, while the scattered components are treated following \citet{Dermer_Menon} and $\gamma$-$\gamma$ absorption in the BLR has been added following \citet{Dermer_Menon}. 

The emission region is assumed to be a relativistic plasma blob of radius $R_{src}$ that moves along the jet with bulk Lorentz factor $\Gamma$, carrying a tangled magnetic field (of intensity $B$) and relativistic electrons with a distribution that steepens with energies, represented here by a broken-power law (BPL) distribution in the range of [$\gamma_{min}$,  $\gamma_{max}$] with a break energy $\gamma_b$ and a sharp cut-off at high energy.  The jet is oriented at a small angle $\theta$ with respect to the line of sight resulting in a Doppler boosting $\delta$.

The accretion disc is described here as a multi-temperature black body spectrum peaking in the optical/UV\footnote{a geometrically thin optically thick accretion disc is considered (\cite{Shakura_Sunyaev_1973})}, while the hot dust in the molecular torus is a simple black body peaking in the near-infrared.  The X-ray corona, above and below the inner parts of the accretion disc, is treated as a simple power law function with an exponential cut-off at around 150 keV (\citet{Ghisellini_Tavecchio_2009}). Accretion disc and corona photons will ionize the BLR, which is assumed to be  a spherical shell of width $\Delta R$ expanding between an inner and an outer radius ($\Delta R \ll R_{in}, R_{out}$).  We assume the BLR clouds intercept and reprocess, mainly into emission lines, a fraction of the disc luminosity (see \citet{Cerruti_2013} for more details). The density of the BLR has a power law shape within $\Delta R$, with an index $\xi$ fixed to $\xi = -2$ that implies that most of the ionization takes place close to the inner edge of the BLR (see \citet{Dermer_Menon}, \citet{Finke_Dermer2010} for a detailed explanation). 

The torus, the corona and the BLR luminosities are scaled to the disc luminosity through the factors $\tau_{IR}$, $\tau_X$ and $\tau_{BLR}$ respectively that are considered as free parameters. These low-energy external radiation fields are comptonized when they encounter the relativistic electrons from the blob {(\citet{Dermer_Menon})}. We consider direct interactions between the relativistic electrons and the torus, the disc and the corona, as well as interactions between BLR-scattered disc and corona photons and the blob.

The effect of both intrinsic ($\gamma$-$\gamma$ opacity due to the BLR radiation field) and extragalactic background light absorption (\cite{Franceschini_2008}) are taken into account. 
A list of the model parameters can be found in Table \ref{tab:model_parameters} and Table \ref{tab:external_parameters} and a detailed description of the code in \citet{Arrieta_Lobo_2018}.

\begin{table}
	\centering
	\caption{Blob model parameters}
	\label{tab:model_parameters}
	\begin{tabular}{|l|  l|}
		\hline
		Parameter & Symbol \\
		\hline
		Bulk Doppler factor											& $\delta$ \\
		Viewing angle                                               & $\theta$\\
		Radius of emitting region   								& R$_{src}$\\
        Magnetic field intensity    								& $B$\\
		Normalization at $\gamma=1$									& $K$\\
		Low-energy spectral index		                            & $n_1$\\ 
		High-energy spectral index                               & $n_2$\\ 
		Minimum electron Lorentz factor								& $\gamma_{min}$\\
		Break electron Lorentz factor									& $\gamma_{b}$\\
		Maximum electron Lorentz factor								& $\gamma_{max}$\\
		Jet height                                                  & R$\gamma$\\
		\hline
	\end{tabular}
\end{table} 

\begin{table}
	\centering
	\caption{External fields model parameters}
	\label{tab:external_parameters}
	\begin{tabular}{|l|  l|}
		\hline
		Parameter & Symbol \\
		\hline
		Disc Luminosity 											& L$_d$ \\
		Black hole mass							                	& M$_{BH}$\\
        Accretion efficiency            							& $\eta$=1/12\\
        Corona lum./ Disc Lum.					                    & $\tau_X$\\
        Coronal X-rays index            							& $\alpha_X$\\
		BLR inner radius						                    & R$_{BLR}$\\
		Dust torus temperature		                                & T$_{IR}$\\ 
		Dust torus radius		                                    & R$_{IR}$\\ 
		Fraction of disc lum. reprocessed by the BLR				& $\tau_{BLR}$\\
		Fraction of disc lum. reprocessed by the torus				& $\tau_{IR}$\\
		Fraction of corona emission reflected by the BLR 			& $\tau_{X,BLR}$=0.01\\
		\hline
	\end{tabular}
\end{table} 

In what follows, two different scenarios regarding the position of the blob ($R_{\gamma}$) are studied and compared to investigate the origin of the HE emission and its variability. 

In the first scenario, disc\&BLR-IC, the disc and BLR inverse Compton components (IC) dominate at hard X-rays and $\gamma$-ray energies respectively ($R_{\gamma} < R_{BLR,in}$).  The photons from the BLR are boosted in the jet frame and consequently the inverse Compton scattering of the BLR photons dominates the HE part of the SED. Since the accretion disc photons are hotter than the ones from the BLR, the disc IC component can still strongly contribute to the radiative output if the blob is close enough to the disc \citep{Dermer_Schlickeiser_1993, Sikora_1994}. In this scenario, we fix $\tau_{IR}$ to 0.05. In the second scenario, torus-IC, the torus IC component dominates at $\gamma$-ray energies ($R_{\gamma} \sim R_{BLR,out}$). The photons from the disc and the BLR are de-boosted in the blob frame and $\tau_{BLR} = 10^{-4}$, so that the dominant contribution to external  processes comes from the dusty torus. 

\section{Physical constraints} \label{sect_physical_constraints}
One of the main difficulties of the broadband SED modelling consists in a high number of free parameters. To reduce the number of free parameters we can rely on observational constraints such as the variability timescales in different frequency bands, and the measurements of the jet opening angles (when available) in order to constrain the radius of the emitting region.

The causality condition puts an upper limit to the intrinsic radius of the source: $R_{src} < c \tau \delta / (1+z)$, where $\tau$ is the variability timescale in a given frequency band and $\delta$ is the Doppler factor. Although variability timescales vary from one flaring episode to another, depending also on the frequency range and the convention used to define them (flux doubling/halving time, total duration etc.), they remain short, of the order of days or hours (especially in $\gamma$-rays). All measurements indicate compact regions in the jet to be responsible for variability in $\gamma$-rays and sometimes in X-rays and Optical/UV bands too.

As usual, we assume that $\Gamma \alpha_{int} \lesssim 1 $, where $\Gamma$ is the bulk Lorentz factor and $\alpha_{int}$ is the intrinsic opening angle of the jet. Assuming a conical jet structure, it also requires the size $R_{src}$ of the emission region to be less than the size of the jet at the emission region location $R_\gamma$, and so $R_{src} \lesssim R_\gamma / \Gamma$.

An even modest change in $\theta$ can be reflected in large changes of $\delta$ and then of the observed flux.  Intermediate values of a few degrees were chosen for the viewing angles. We could exclude much larger or smaller values due to the (de-)boosting effect, causing the model to strongly underestimate or overestimate the overall flux. For example $\sim$ 10$\degree$ and 0.5$\degree$ were estimated for 1H\,0323$+$342  and PMN\,J0948$+$0022, respectively, by \cite{Homan_2021_Mojave}, based on MOJAVE program data, \citep{Lister_2021}, which we cannot reproduce for the modelled flux states.

The knowledge of the black hole mass and of the luminosity of the accretion disc considerably help
to constrain the free parameters of the model. We thus make use of the estimates available in the literature and restrict our scenarios to sub-Eddington accretion regimes. The inner radius of the BLR, $R_{BLR,in}$, is estimated from the empirical relations derived from reverberation mapping campaigns (e.g. \citet{Bentz_2013}, \citet{Kaspi_2005}) as $R_{BLR,in}\simeq 10^{17}L_{d,45}^{1/2}$ [cm] and we adopt $R_{BLR,out} = 3R_{BLR,in}$.  Scattering of dust torus photons occur at larger  distance $R_{IR} = 3.5\times10^{18} L_{d,45}^{1/2}(\frac{T_{IR}}{10^3})^{-2.6}$ [cm] (e.g. \cite{Sikora_2009}), where $T_{IR}\sim 1200-1500 K$ to avoid sublimation of the dust grains.
$L_{d,45}$ refers to the disc luminosity in units of $10^{45}\ \rm{erg\ s^{-1}}$.

\section{1H 0323+342} \label{sect_1H}
\subsection{Observational constraints and physical properties}
1H 0323+342 is the closest source among currently known $\gamma$-NLS1 galaxies (z=0.0625,  \citet{Landt_2017}). 

1H\,0323+342 is thought to be moderately radio-loud, with radio-loudness (RL) ranging from  4 to 25 (\citet{Zhou_2007}). 
 
\cite{Fuhrmann_2016} characterized the pc-scale jet structure and kinematics of 1H\,0323+342 using multi-epoch radio VLBA observations (MOJAVE\footnote{Monitoring of Jets in Active galactic nuclei with VLBA Experiments} programme, \cite{Lister_2009}), complemented with single-dish Effelsberg and IRAM data.  In addition to the compact stationary core, six moving knots were identified, among which five were found to have superluminal apparent velocities $\sim$ (1-7)\,c, suggesting relatively low viewing angle values, $\theta \leq$ 4\degree - 13\degree. If we make the reasonable assumption of $\theta \sim$ 1/$\Gamma$, this constrains $\Gamma$ > 4.4–14.3. Multi-frequency radio light-curves, from 2010 to 2015, show a major outburst in early 2013, followed by the detection of a new radio component around 2013.5. In  a more recent work \citet{Homan_2021_Mojave} characterized the pc-scale jet of a wide
sample of AGN from MOJAVE program for kinematics \citep{Lister_2021}, including 1H\,0323+342, and, assuming the  median brightness temperature of their sample derive  $\Gamma \sim$ 11 and $\theta \sim$ 9.8\degree for the median state.

\citet{Foschini_2009} found remarkable variations in the X-ray emission above 20\,keV: the source went from a low flux, soft spectrum state in 2004 to a high flux, hard spectrum in 2006-2008, which can be interpreted as disc/corona emission dominating in 2004 and a more important jet contribution in 2006-2008. \citet{Yao_2015} showed that the source presents statistically correlated variability in Ultra-Violet and X-ray bands, both on daily and yearly timescales, the correlated UV/X-ray variability timescale of days being consistent with X-ray emission from reprocessed accretion disc photons. 

From long-term follow-up of the $\gamma$-ray emitting NLSy1 galaxies by \textit{Swift} in Optical, UV and X-ray bands, \cite{D'Ammando_2020_Swift_view} reported very short X-ray variability timescale for 1H 0323+342 in July 2013, $\tau = 6\ \rm{ks}$ hence $R_{src} < 1.8 \times 10^{15}$ cm assuming a typical value of $\delta=10$. A systematic analysis of all \textit{Swift}-UVOT observations in optical and UV bands of this source up to April 2019 presented in \cite{D'Ammando_2020_systematic_analysis_swift} showed a very short intrinsic variability timescale in UV too, $\sim$ 5.4 to 16.3 ks for 1H\,0323+342, implying $R_{src} < (1.6 - 4.9) \times 10^{15}$ cm.  \cite{Paliya_2015} reports a flux-doubling time scale of $3.09 \pm 0.85$ hours during the \textit{Fermi}-LAT flare of August 2013 ($\sim$ 56531 MJD). This value corresponds to $R_{src} \lesssim (2 - 4) \times 10^{15}$ cm. All X-ray, Optical/UV and $\gamma$-ray variability timescales agree. 

For all three sources, to assemble (quasi-)simultaneous multiwavelength data sets depicting different activity states, we took advantage of the \textit{Fermi} All-sky Variability Analysis web tool\footnote{\url{https://fermi.gsfc.nasa.gov/ssc/data/access/lat/FAVA/}} (FAVA), a tool that blindly searches for transients over the entire sky observed by the LAT (Abdollahi et al. 2017) and retrieved data from the {\it Fermi}-LAT Light Curve Repository\footnote{\url{https://fermi.gsfc.nasa.gov/ssc/data/access/lat/LightCurveRepository/about.html}} (\cite{Kocevski_2021ATel}). The flaring and quiescent states identified in HE were then cross-matched with \textit{Swift} X-rays observations. 

For most of the time, the $\gamma$-ray flux did not much exceed the value averaged over time (see Fig.\ref{fig:1H_multifreq_LC}). The only prominent flare occurred  in August 2013, featuring a much harder spectrum (\citet{Sahakyan_2018}). During this flare, the flux-doubling time was as short as $0.49 \pm 0.17 $ day (\citet{Baghmanyan_2018}), implying the emission is produced in a compact region, $R_{src} \lesssim 1.2 \times 10^{15}$ cm assuming $\delta$ = 10. 

The black hole mass of 1H\,0323+342 was estimated to be $(1.5-2.2)\times10^7\sunmass$ by \cite{Landt_2017} using H$\alpha$, H$\beta$ and Pa$\alpha$ line width and continuum luminosities. It is consistent with the value of $(1.8-3)\times10^7\sunmass$ obtained by \cite{Zhou_2007}. Short X-ray variability timescales from a deep \textit{Suzaku} observation also suggest a relatively small mass of the BH, of $M_{BH} \sim 10^7\sunmass$ \citep{Yao_2015}.

\citet{Kynoch_2017} observationally constrained the external photon fields of 1H\,0323$+$342 through a quasi-simultaneous near-IR, optical and X-ray dataset, allowing to narrow the $L/L_{Edd}$ range to 0.6-0.8 for a non-rotating Schwarzschild BH.  Very high spin values and super-Eddington case were rejected because of the poor quality of the fit to the data in that case. The luminosity of the accretion disc was estimated to be $2.1\times10^{45}\,\rm{erg\,s^{-1}}$. 

For our study, we consider three activity states (low, intermediate and high) and investigate the evolution of the jet power of 1H\,0323$+$342 (see Table \ref{tab:1H0323_obs}).

\subsection{The dataset}

The low activity state SED has been reported in \cite{Kynoch_2017}. The \textit{Fermi}-LAT low state from August to September 2015 is indeed a non flaring period (see Fig.\ref{fig:1H_multifreq_LC}). Contrary to the general case where low states observed by \textit{Fermi}-LAT are rather intermediate states since $\gamma$-NLS1 galaxies tend to be undetected during very low flux phases, 1H~0323+342 went through a genuinely low activity phase during August – September 2015. Quasi-simultaneous GNIRS IR spectral data retrieved from \citet{Kynoch_2017} are added for completeness in the SEDs.
The HE data are complemented with the optical and X-ray data closest in time with this very low \textit{Fermi} state: simultaneous \textit{XMM-Newton} (Optical Monitor and EPIC) data. \textit{Swift}-XRT observations reported by \cite{D'Ammando_2020_Swift_view} show one of the lowest X-ray fluxes of 1H\,0323+342 ever measured with \textit{Swift} during this epoch. Only at the very end of the selected \textit{Fermi}-period, the X-ray flux reaches a 
high value. The corresponding data points are depicted in green in Fig.\ref{fig:1H_Disc_BLR} and \ref{fig:1H_Torus}.

We model the flaring state from August 2013 that corresponds to the largest high energy flux measured up to date for 1H~0323+342 (see Fig.\ref{fig:1H_multifreq_LC} and 'F3' flaring state presented in \cite{Paliya_2014}). The multi-wavelength data used to build  the flaring state SED are retrieved from \cite{Paliya_2014} (Table \ref{tab:1H0323_obs}) and depicted in orange (see Fig.\ref{fig:1H_Disc_BLR} and \ref{fig:1H_Torus}). \citet{D'Ammando_2020_Swift_view} reports the analysis of all \textit{Swift} observations available up to April 2019 and show 1H\,0323+342 to be at maximum in X-rays and optical during this major $\gamma$-ray flare, with a significantly hardened spectrum.

The quiescent state of 2008, modelled by \citet{Paliya_2014}, turns out to rather be an intermediate HE state (Fig.\ref{fig:1H_multifreq_LC}), but it corresponds to very low \textit{Swift}-XRT and UVOT flux levels, among the lowest ever measured for 1H\,0323+342.  The corresponding data points are depicted in yellow in Fig.\ref{fig:1H_Disc_BLR} and \ref{fig:1H_Torus}.

For completeness, we also included in the SEDs NuSTAR observations carried out during 2014 March 15-18 (MJD 56731 - 56733) (\cite{Landt_2017}) along with the contemporaneous Effelsberg radio data taken during 2014 March 01-11 (MJD 56717-56727) (included in the F-Gamma program (\citet{Angelakis_2015})), plotted in blue. Although not simultaneous with the selected low and high states, NuSTAR data are particularly interesting for the modelling purposes since they cover the hard X-ray part of the spectrum as do the catalogue data from the \textit{Swift}-BAT 105-month all-sky survey\footnote{\url{https://swift.gsfc.nasa.gov/results/bs105mon/} }\citep{Oh_BAT_2018}.

Archival data points are extracted from SSDC database\footnote{\url{https://www.ssdc.asi.it}} and depicted in gray.

\begin{table}
	\centering
	\caption{Quasi-simultaneous observations of 1H 0323+342}
	\label{tab:1H0323_obs}
	\begin{tabular}{lcc} 
	    
		Instrument & Time period & MJD \\
		\hline
		\multicolumn{3}{l}{Low state} \\
		\hline
		\textit{Fermi}-LAT & 1 Aug - 30 Sep 2015 & 57235 - 57295 \\ 
		\textit{XMM-Newton} & 23 Aug 2015 & 57257 \\
		GNIRS & 16 Sep 2015 & 57281 \\
		\hline
		\multicolumn{3}{l}{Intermediate state} \\
		\hline
		\textit{Fermi}-LAT & 5 Nov - 5 Dec 2008 & 54775 - 54805 \\ 
		\textit{\textit{Swift}} & 16 Nov 2008 & 54786 \\
		\hline
		\multicolumn{3}{l}{High state} \\
		\hline
        \textit{Fermi}-LAT & 27 - 31 Aug 2013 & 56531 - 56535 \\
        \textit{\textit{Swift}} & 30 Aug 2013 &  56534 \\
		\hline
	\end{tabular}
\end{table}

\subsection{SED modelling}

In the following, we use $L_d = 2\times10^{45}\ \rm{erg\ s^{-1}}$, $M_{BH} = 2\times10^7\ \sunmass$ and $\theta = 5 \degree$, corresponding to the lower limit determined by \cite{Fuhrmann_2016}. We could exclude larger values due to deboosting effect, causing the model to strongly underestimate the overall flux.
The Eddington ratio $l_{Edd}$\footnote{defined as the ratio of the disc luminosity $L_d$ over the Eddington luminosity $L_{Edd} = 1.3\times10^{38}\frac{M_{BH}}{\sunmass}$ $\rm{erg\ s^{-1}}$ } is then 0.80. The distance of the BLR from the central BH, $R_{BLR, in}$, constrained by the disc luminosity, is set to 4.77$\times10^{4}\, \rm{R_{G}}$.  

\begin{figure*}
     \centering
     \begin{subfigure}
         \centering
         \includegraphics[width=0.7\linewidth]{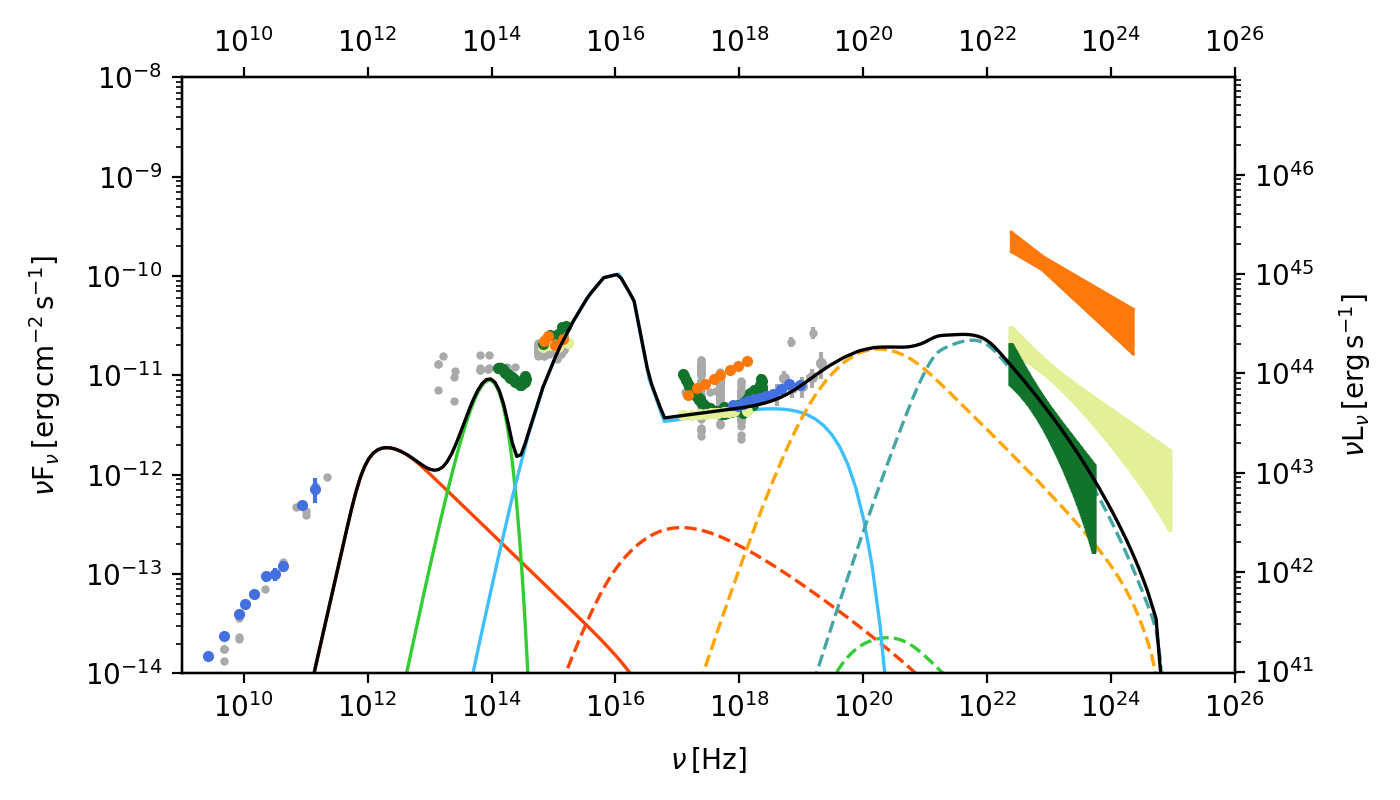}
     \end{subfigure}
     
    \begin{subfigure}
         \centering
         \includegraphics[width=0.7\linewidth]{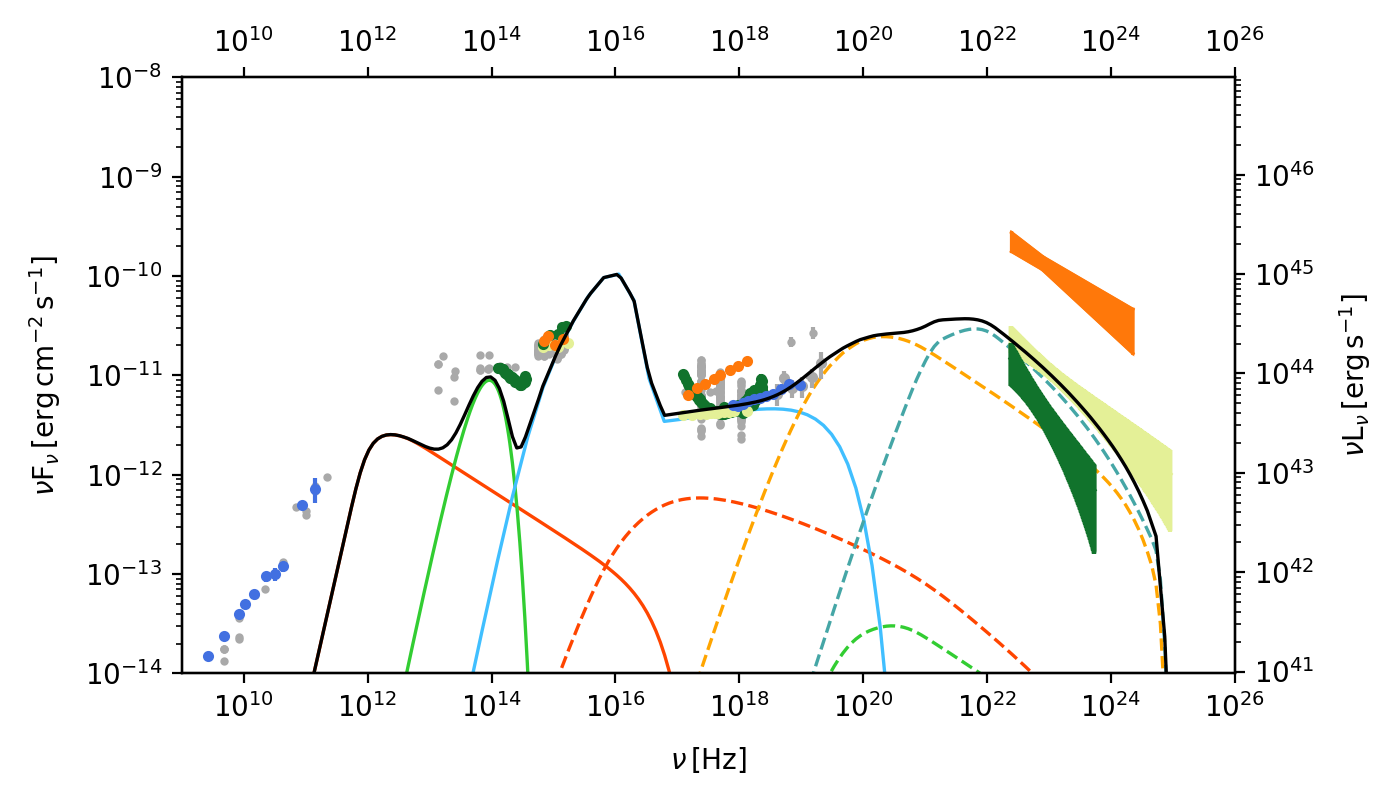}
     \end{subfigure}
     
     \begin{subfigure}
         \centering
         \includegraphics[width=0.7\textwidth]{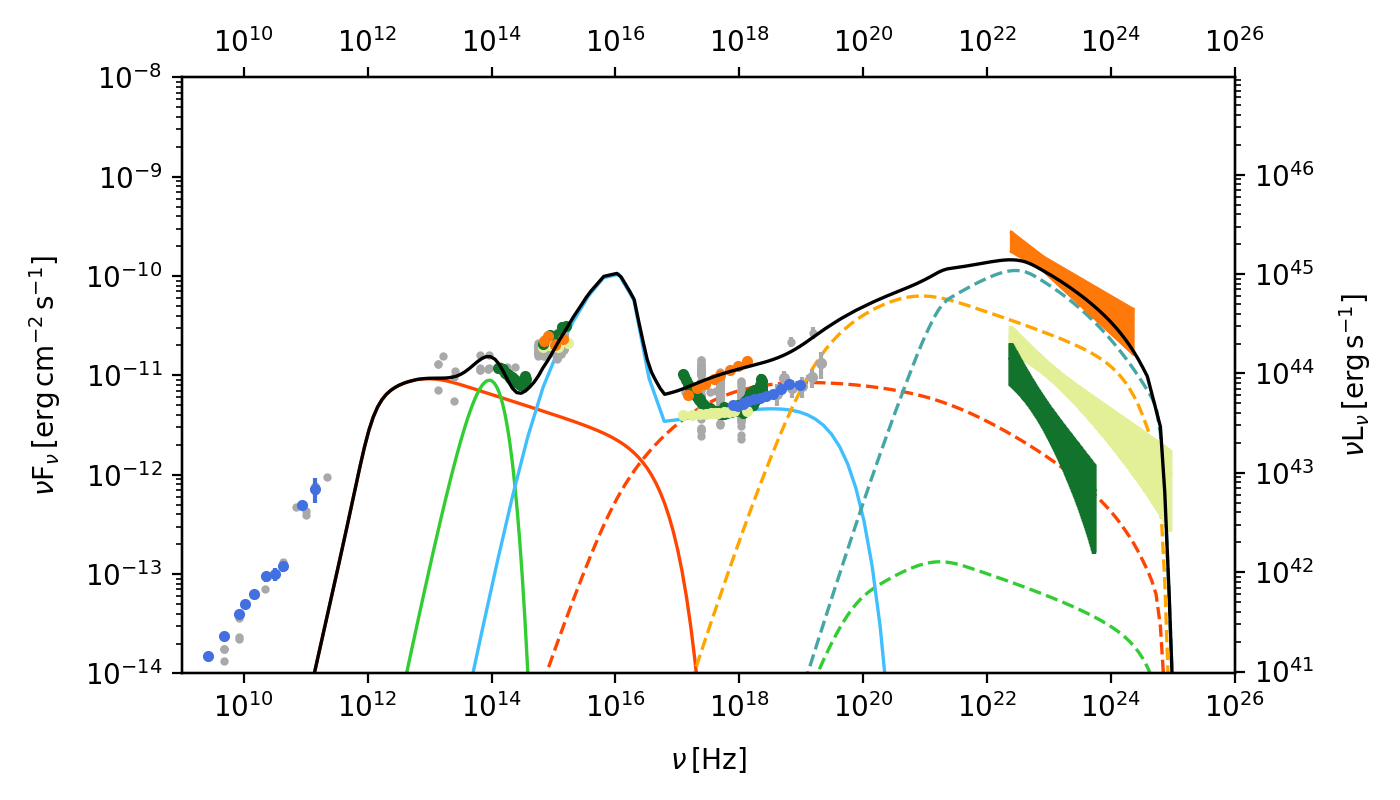}
     \end{subfigure}
        \caption{Disc\&BLR dominated scenario for 1H 0323+342. Direct emission components are presented in solid lines and the associated reprocessed components are depicted in dashed lines of the same color (red for synchrotron and SSC emission, light green for direct torus emission and EIC-torus component, light blue for the direct emission of the accretion disc and corona, whereas the dashed light blue lines represent the IC scattering of the BLR photons by energetic electrons of the blob. Orange dashed lines refer to the emission produced by the IC-scattering of the accretion disc photons. The data points presented here (archival and quasi-simultaneous) are dereddened and deabsorbed. See text for more details.}
        \label{fig:1H_Disc_BLR}
\end{figure*}

For three activity states, two scenarios, disc and BLR-dominated (disc\&BLR) and torus-dominated, are tested and reported in Fig. \ref{fig:1H_Disc_BLR} and \ref{fig:1H_Torus}, respectively. The model changes from lower to higher activity state allowing only the compact jet parameters to vary, while all parameters linked to the external photon fields remain constant. Input parameters as well as physical quantities derived from the model are shown in table \ref{pars_table_1H}. 

The electron distribution is strongly constrained by the \textit{Fermi}-LAT spectral shape. The clear hardening of the GeV spectrum during the flaring state, compared to the quiescent ones, imposes a hardening of the electron distribution. The transition from a quiescent to a more energetic state is simply described by a denser and more energetic blob within the jet (larger  electron density and Doppler factor). 

In the disc\&BLR scenario, reasonable account of the SEDs is achieved with an emitting region located at $R_\gamma = 2.5\times10^3\ [\rm{R_G}]$. The relativistic jet peaks at sub-millimeter frequencies. \citet{Kynoch_2017} attributed the IR part of the SED, as defined by the Spitzer and the WISE data, to thermal emission from the extended dusty torus, thus requiring that the synchrotron emission does not contribute substantially to it. In the highest activity state this may not be fulfilled, but those IR data are not simultaneous nor contemporaneous and for an even more powerful jet, one might have a dominating synchrotron component. The thermal disc emission is a major contribution at optical/UV wavelengths. Hard X-ray data are ascribed to a combination of the direct corona emission and jet radiation, the contribution from SSC radiation dominating in the flaring state. 

In the second scenario, the blob is located at the outermost edge of the BLR. A much larger torus reprocessing factor is needed here, $\tau_{IR}$ = 0.25, to account for the high energy emission. The large fraction of disc reprocessed by the dusty torus makes its direct emission to peak higher than the archival data regardless of the activity state.  The peak of the electron distribution occurs at much higher energies in the torus-dominated case, so all IC components of the model are shifted to higher energies. The electron energy densities are much lower in the torus-dominated scenario, which implies less synchrotron self-absorption at radio frequencies.  In this scenario the size of the emission region is in contradiction with the variability constraint we have adopted. 

In both scenarios, our model underestimates the UV and X-ray radiation during the quiescent state. This could be ascribed to our simple description of the accretion flow at these frequencies as a multi-temperature blackbody.

\begin{table*}
\caption{Model parameters for the different scenarios considered for 1H 0323+342. \\
Input parameters: (a) redshift, (b) viewing angle, (c) BH mass, (d) gravitational radius size, (e) disc luminosity, (f) Eddington ratio, (g) mass accretion rate, (h) Doppler factor of the blob, (i) particle density normalization (at $\gamma = 1$),  (j) size of the emitting region, (k) magnetic field, (l) and (m) first and second indices of the broken power law particle distribution, (n), (o) and (p) minumum, break and maximum electron energies, respectively, (q) dusty torus temperature, (r) reprocessing factor of the torus, (s) and (t) slope and reprocessing factor of the corona, (u) blob-BH distance, (v) and (w) reprocessing factor and inner radius of the BLR. \\ Derived parameters: (1) ratio of electron distribution kinetic energy density and magnetic field energy density, (2) particle density, (3) electron energy density, (d) total jet power.}
    \label{pars_table_1H}
\begin{minipage}{\textwidth} 
    \centering
    \begin{tabular}{lcccccc}
    \hline
    Source description & \multicolumn{6}{c}{} \\
    \hline
    $z\ ^{a}$ & \multicolumn{6}{c}{0.0625} \\
    $\theta\, (\degree)\ ^{b} $ & \multicolumn{6}{c}{5} \\
    $M_{BH}\, [\rm{\sunmass]}\ ^{c}$ & \multicolumn{6}{c}{$2\times10^7$} \\
    $R_G\, [\rm{cm}]\ ^{d}$ & \multicolumn{6}{c}{$2.95\times10^{12}$} \\
    $L_{Disc}\, [\rm{erg\ s^{-1}}]\ ^{e}$ & \multicolumn{6}{c}{$2\times10^{45}$} \\
    $l_{Edd}\ ^{f} $ & \multicolumn{6}{c}{0.79}\\
    $\dot m\, [\rm{\sunmass/yr}]\ ^{g}$ & \multicolumn{6}{c}{0.42} \\
    \hline

    Scenario & \multicolumn{3}{c}{Disc \& BLR} & \multicolumn{3}{c}{Torus} \\
    \cline{1-7}
    State & Low & Intermediate & Flare & Low & Intermediate & Flare \\
    \hline
    
    $\delta\ ^{h}$ & 9 & 9 & 10 & 9 & 9 & 11 \\
    $K\, [\rm{1/cm^3}]\ ^{i}$ & $4\times10^6$ & $5\times10^6$ & $8.4\times10^6 $ & $7.4\times10^6$ & $5.0\times10^{6}$ & $1.3\times10^7$ \\
    $R_{src}\, [\rm{cm}]\ ^{j}$ & $1.15\times10^{15}$ & $1.15\times10^{15}$ & $1.01\times10^{15}$ & $1.40\times10^{16}$ & $1.43\times10^{16}$ & $1.34\times10^{16}$ \\
    $B\, [\rm{G}]\ ^{k}$ & 2.3 & 2.3 & 2.3 & 0.30 & 0.30 & 0.30 \\
    $n_1\ ^{l}$ & 2.2 & 2.2 & 2.2 & 2.7 & 2.7 & 2.7 \\
    $n_2\ ^{m}$ & 4.2 & 3.8 & 3.4 & 4.7 & 4.0 & 3.9 \\
    $\gamma_{min}\ ^{n}$ & 50 & 50 & 50 & 500 & 500 & 500 \\
    $\gamma_{b}\ ^{o}$ & 150 & 150 & 280 & 600 & 650 & 680 \\
    $\gamma_{max}\ ^{p}$ & $2\times10^4$ & $2\times10^4$ & $2\times10^4$ & $4\times10^4$ & $4\times10^4$ & $4\times10^4$ \\
    \hline
    
    $T_{IR}\, [\rm{K}]\ ^{q}$ & 1200 & 1200 & 1200 & 1200 & 1200 & 1200 \\
    $\tau_{IR}\ ^{r}$ & 0.05 & 0.05 & 0.05 & 0.25 & 0.25 & 0.25\\
    $\alpha_X\ ^{s}$ & 0.9 & 0.9 & 0.9 & 0.9 & 0.9 & 0.9 \\
    $\tau_X\ ^{t}$ & 0.15 & 0.15 & 0.15 & 0.17  & 0.17 & 0.17 \\
    $R_\gamma\, [\rm{R_G}]\ ^{u}$ & $2.5\times10^3$ & $2.5\times10^3$ & $2.5\times10^3 $ & $1.5\times10^5$ & $1.5\times10^5$  & $1.5\times10^5$ \\
    $\tau_{BLR}\ ^{v}$ & 0.28 & 0.28 & 0.28 & $1\times10^{-4}$ & $1\times10^{-4}$ & $1\times10^{-4}$ \\
    $R_{in}^{BLR}\, [\rm{R_G}]\ ^{w}$ & $4.77\times10^4$ & $4.77\times10^4$ & $4.77\times10^4$ & $4.77\times10^4$  &  $4.77\times10^4$ & $4.77\times10^4$ \\ 
    \hline

    $u_e/u_b\ ^{1}$ & 9.72 & 12.87 & 29.54 & 10.95 & 9.75 & 27.14 \\

    $n_e\, [\rm{1/cm^3}]\ ^{2}$ & $2.57\times10^4$ & $3.27\times10^{4}$ & $6.07\times10^4$ & 68.56 & 55.44 & $1.51\times10^2$  \\

    $u_e\, [\rm{erg\, cm^{-3}}]\ ^{3}$ & 2.05 & 2.71 & 6.22 & $3.92\times10^{-2}$ & $3.49\times10^{-2}$ & $9.72\times10^{-2}$ \\

    $P_{jet,tot}\, [\rm{erg\, s^{-1}}]\ ^{4}$ & $3.82\times10^{44}$ & $5.02\times10^{44}$ & $1.40\times10^{45}$ & $2.39\times10^{44}$ & $2.28\times10^{44}$ & $1.63\times10^{45}$ \\

    \hline
    \end{tabular}
    \end{minipage}
\end{table*}

\section{PMN\,J0948+0022} \label{sect_PMN}

\subsection{Observational constraints and physical properties} \label{PMN_obs_constraints}
PMN\,J0948+0022, with a moderately high redshift (z = 0.5846, \cite{Zhou_2003}), is classified as a typical radio loud NLSy1 galaxy with strong ($RL \gtrsim 10^{3}$) and variable radio emission. Its inverted radio spectrum and very high brightness temperature ($\simeq 10^{13}\ \rm{K}$) derived from variation of the radio flux
suggested the presence of a relativistic jet strongly beamed towards the observer (\cite{Zhou_2003}). This was later confirmed when PMN\,J0948+0022 turned out to be  the first narrow-line Seyfert 1 galaxy detected in $\gamma$-rays by the \textit{Fermi}-Large Area Telescope (\cite{Abdo_2009_Discovery}).

VLBI monitoring during the first $\gamma$-ray flare ever detected in a NLS1, that occurred in July 2010, with shortest and even intraday variability time scales (\citet{Foschini_2011}) revealed polarization, change of the EVPA vector of about 90\degree \, and a variable radio spectral shape, probably related to the $\gamma$-ray flare. The source is seen as a bright compact core, with large brightness temperatures (\cite{Giroletti_2011}), indicating a heavy Doppler beaming. Since then its variable radio spectrum has been intensively monitored (see e.g \cite{Angelakis_2015} and \cite{Laehteenmaeki_2017} for a review). The existence of several superluminal components in the jet of this source points towards a small viewing angle and large Lorentz factor (e.g. \cite{Homan_2021_Mojave}).

Although PMN J0948+0022 already showed a moderately enhanced $\gamma$-ray emission during March-July 2009 (\cite{Abdo_2009_MWL_campaign}), the outburst from July 2010 was a factor of $\sim 2$ greater, exceeding a peak flux value of $\sim 10^{-6}\, \rm{ph \, cm^{-2}\, s^{-1}}$ (corresponding to an observed luminosity of $\sim 10^{48}\ \rm{erg\ s^{-1}}$) in the 0.1-100 GeV energy range, for the first time in this type of source. Long term multi-wavelength (MWL) monitoring from \cite{Foschini_2012} showed that the flare from July 2010 is accompanied by a more progressive flaring episode detected in radio wavebands (at 15 GHz and 86 GHz). 
The source showed prolonged activity in radio in 2011 as well as in $\gamma$-rays. 

From a systematic analysis of about 10 years of \textit{Swift} observations of NLSy1 galaxies  \cite{D'Ammando_2020_systematic_analysis_swift} reported rapid variability in optical bands for PMN J0948+0022, ranging from very short (minimum) intrinsic variability timescale of $\sim$ 3.3 ks to $\sim$ 30 ks, that translates into $R_{src} < (1.1-8.8) \times 10^{15}$ cm, assuming a Doppler factor of 10. Similarly, $\gamma$-ray variability timescale values $74.6 \pm 27.6$ hours (\cite{Paliya_2015}), $2.3 \pm 0.5$ days (\citet{Foschini_2012})  and 3-30 days reported by \cite{Calderone_2011}, correspond to $R_{src} < (3 - 7) \times 10^{16}$ cm, $R_{src} < (3 - 5) \times 10^{16}$ cm and $(0.5 - 6) \times 10^{17}$ cm respectively, assuming a Doppler factor of 10. \citet{Sahakyan_2018} even quote $18.9$ hours for the extremely active state  during the  period 2012 December - 2013 January. The large dispersion in the estimations is due to different variability trends in different spectral domains and binning during extreme flaring activity episodes compared to intermediate/quiescent ones. Indeed  PMN J0948+0022 is a very active $\gamma$-ray emitter, the source is alternatively in its flaring or quiescent states (see Fig.\ref{fig:PMN_multifreq_LC}). We keep the most restrictive upper limit $R_{src} \lesssim (1.1-8.8) \times 10^{15}$ cm.

A lower limit on the BH mass of $10^8 \sunmass$ was obtained from accretion disc fitting of \textit{Swift}-UVOT data  (\cite{Abdo_2009_Discovery}, \cite{Calderone_2013}), in agreement with the BH mass range obtained by \cite{Zhou_2003}: $M_{BH} \sim (4.0\times10^7 - 8.1\times10^8) \sunmass$ deduced with the assumption that the line emitting cloud motions are virialized.  The estimate from \cite{Yuan_2008} $\sim10^7 \sunmass$ remains significantly smaller and inconsistent with the previous values. More generally, \cite{Chiaberge_Marconi_2011} found that taking into account radiation pressure correction on the BLR clouds in the virial techniques results in higher BH masses, which illustrates the complexity of BH mass measurements. High radio-loudness parameter seems to favor larger BH masses too. Since different independent techniques seem to converge towards $M_{BH} \gtrsim 10^8\ \sunmass$, we keep $M_{BH} = 1.5\times 10^8 \sunmass$ as assumed for modelling in \cite{Abdo_2009_Discovery}. 

Disc luminosity has been estimated to $L_{d} = 9\times10^{45}\, \rm{erg\ s^{-1}}$ obtained from SED fitting of the 2009 state (see \cite{Abdo_2009_new_class_of_GR_AGN}, \cite{Abdo_2009_MWL_campaign} for more details), and $L_{d} = 5.7\times10^{45}\ \rm{erg\ s^{-1}}$ from modelling of the 2011 and 2013 states in \cite{D'Ammando_2015}. We used $L_{d} = 9\times10^{45}\ \rm{erg\ s^{-1}}$ because lower values did not provide enough disc luminosity for EIC scattering of the BLR and torus photons to account for the observed elevated HE flux levels.

\subsection{The dataset}

\textit{Fermi}-LAT detected the most powerful flaring activity from PMN\,J0948$+$0022 between 2012 December and 2013 January, triggering \textit{Swift}, VERITAS\footnote{Very Energetic Radiation Imaging Telescope Array System;  \url{https://veritas.sao.arizona.edu/}} and other MWL observations \citep{D'Ammando_2015}. VERITAS observations resulted in upper limits. A strong optical/UV and X-ray flare occurred quasi-simultaneously with the $\gamma$-ray flare, implying that the jet emission is the dominant mechanism. One of the most prominent radio flares from this source at 15 Ghz (\cite{Lahteenmaki_2017}) is recorded a few days later. Bulk Lorentz factors as high as 30 were used by \citet{D'Ammando_2015} to model the flare SED, which is not surprising given the large observed luminosities and  BL Lac-type shape, with a completely jet-dominated SED  at that time. Being interested in the disc dominated cases, we discard the 2013 flaring event from our analysis.

In a compromise between photon statistics (see Fig. \ref{fig:PMN_multifreq_LC}) and reasonable multi-wavelength coverage, we consider, for the low state, the dataset from \citet{D'Ammando_2015} that features \textit{Fermi}-LAT data from 22 May to 11 June, \textit{XMM-Newton} on  28-29 May 2011  and Effelsberg radio data at 15 and 32 GHz on 24 May 2011, avoiding nearby HE outburst phases (especially on 21 June 2011, see e.g. \citet{Foschini_2012}). This state is particularly interesting since the soft-excess is clearly visible in the X-ray data, and a hard power-law spectrum describes the X-ray data beyond 2.5keV (\citet{Bhattacharyya_2014}, \cite{dammando_2014}). This contemporaneous data set is depicted in green in Figures \ref{fig:PMN_Disc_BLR} and \ref{fig:PMN_Torus}.

The most prominent $\gamma$-ray outburst from July 2010 occurred during the course of a dedicated multi-wavelength campaign performed during July–September 2010 (\cite{Foschini_2011}). Only one X-ray observation close to the outburst could be performed by \textit{Swift} on July 3rd, about 4 days before the $\gamma$-ray outburst. It can be seen from Fig.\ref{fig:B2_multifreq_LC} that the corresponding \textit{Swift}-XRT and UVOT fluxes are indeed only moderately high.  

The dataset for the 2010 flare is taken from the study by \citet{Foschini_2012} as they feature mostly simultaneous or contemporary data. From the 5 different epochs presented in their paper, we have considered that of July 8th, 2010 (MJD 55386), which corresponds to the first documented $\gamma$-ray outburst of the source, as a high state for modelling. For their dataset, \citet{Foschini_2012} considered \textit{Fermi}-LAT data integrated over one day while \textit{Swift} observations from July 3rd are considered. Effelsberg radio data from 18 September 2010 (MJD 55457) retrieved from \cite{Angelakis_2015} are also included. The contemporaneous SED is depicted in orange on Figures \ref{fig:PMN_Disc_BLR} and \ref{fig:PMN_Torus}. Mid infrared photometry for the source was obtained by the Wide-field Infrared Survey Explorer\footnote{\url{https://irsa.ipac.caltech.edu/Missions/wise.html}}  (WISE) during May 14-15 (MJD 55330-55331) and \,November 20-21 (MJD 55520-55521), that is right before and after the period considered here. The average flux per period in gray along with with archival data points extracted from the ASDC database.

To build the SED we analyzed \textit{Fermi}-LAT data recovered from the \textit{Fermi} Science Support data server \footnote{\url{https://fermi.gsfc.nasa.gov/cgi-bin/ssc/LAT/LATDataQuery.cgi}}; details on  the analysis can be found in Appendix \ref{sect_data_analysis}. Assuming a power-law spectral shape, a binned likelihood analysis of the flaring state from July 7-9 (MJD 55384 - 55386) yields a detection with a Test Statistic  TS = 175.7 (i.e. 13.3$\sigma$) in the 0.1-100 GeV energy range, $F_{0.1-100GeV} = (1.07 \pm 0.15) \times 10^{-6} \ \rm{ph \ cm^{-2} \ s^{-1}}$ for  $\Gamma = 2.22 \pm 0.12$.  For the period between May 22 to June 11 2011 (MJD 55703 - 55723), the analysis yielded a detection with  TS = 107.2 (i.e. 10.4$\sigma$) in the 0.1–100 GeV energy range, an integral flux $F_{0.1-100GeV} = (2.12 \pm 0.35) \times 10^{-7}\, \rm{ph\, cm^{-2}\, s^{-1}}$ for a softer spectral slope, $\Gamma  = 2.50 \pm 0.14$, than the high state. The highest energy photon from the source is below 4 GeV.

\begin{table}
	\centering
	\caption{Observations of PMN J0948+002}
	\label{tab:PMNJ0948_obs}
	\begin{tabular}{lcc} 
		Instrument & Time period & MJD \\
		\hline
		\multicolumn{3}{l}{Low/intermediate state} \\
		\hline
		\textit{Fermi}-LAT & 22 May - 11 Jun 2011  & 55703 - 55723 \\
		\textit{XMM-Newton} & 28 - 29 May 2011 & 55709 - 55710 \\
		Effelsberg & 24 May 2011 & 55705 \\
		\hline
		\multicolumn{3}{l}{High state} \\
		\hline
		\textit{Fermi}-LAT & 7 - 9 Jul 2010 & 55383.5 - 55389.5 \\
		\textit{\textit{Swift}} (UVOT \& XRT) & 3 Jul 2010 & 55380 \\
	    Effelsberg & 18 Sep 2010 & 55457 \\
	\end{tabular}
\end{table}

\subsection{SED modelling} \label{PMN_SED_model}
All models assume  a viewing angle of $\theta = 3 \degree$ \citep{Foschini_2011}, a black-hole mass of $M_{BH} = 1.5 \times 10^{8}\, \sunmass$ and a disc luminosity of  $L_{d} = 9\times10^{45} \rm{erg\ s^{-1}}$ corresponding to an Eddington ratio $l_{Edd}$ = 0.48. The distance of the BLR from the central BH, $R_{BLR, in}$, constrained by the disc luminosity, is set to $1.5 \times 10^{4}\,\rm{R_{G}}$.

\begin{figure*}
     \centering
     \begin{subfigure}
         \centering
         \includegraphics[width=0.7\linewidth]{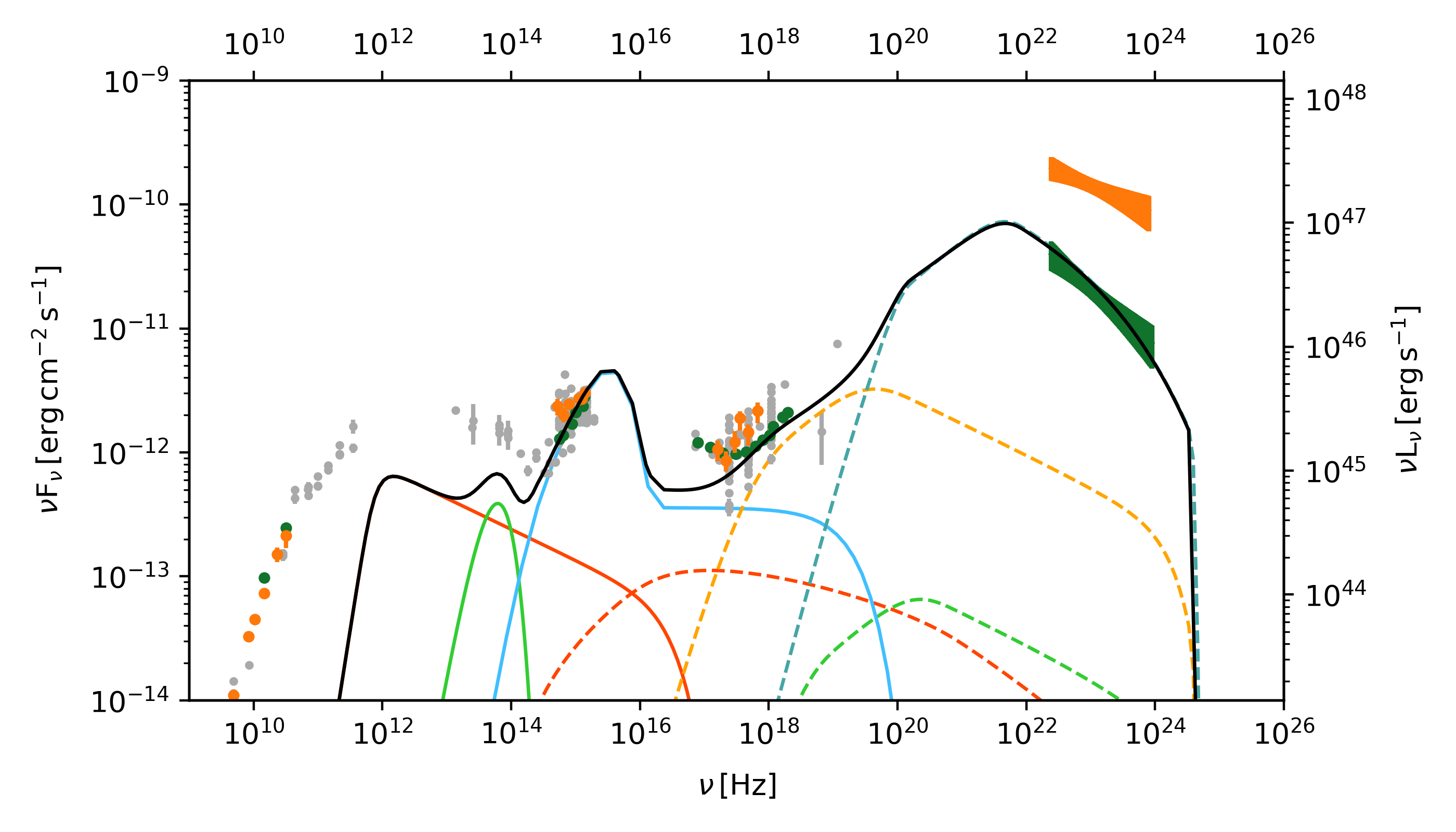}
     \end{subfigure}
     \begin{subfigure}
         \centering
         \includegraphics[width=0.7\textwidth]{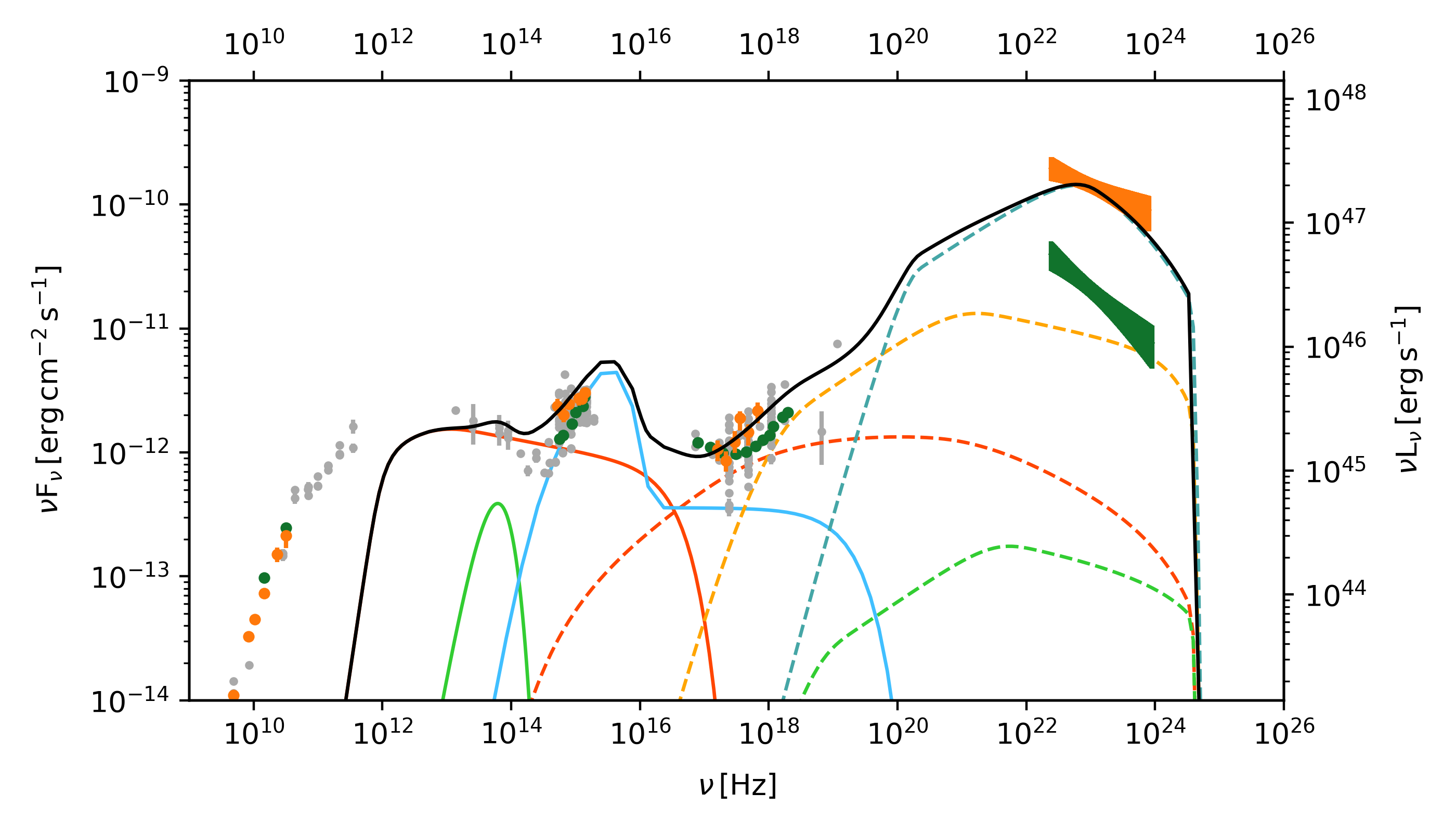}
     \end{subfigure}
        \caption{Disc\&BLR dominated scenario for PMN J0948+0022. The description of the model components and the data is the same as in Fig. \ref{fig:1H_Disc_BLR}.}
        \label{fig:PMN_Disc_BLR}
\end{figure*}

Again, the intermediate and the flaring states of PMN\,J0948$+$0022 were modelled in a disc\&BLR-IC scenario where the blob is below the BLR and in a torus-IC scenario where the emission region is at the outermost edge of the BLR. 
The results of the broad-band modelling of the two activity states for PMN\,J0948+0022 are shown in Fig.\ref{fig:PMN_Disc_BLR} and \ref{fig:PMN_Torus} in the BLR- and torus-dominated scenarios, respectively and the corresponding parameters are given in table \ref{pars_table_PMN}.

In the disc\&BLR dominated scenario, the BLR reprocesses 40\% of the disc luminosity and the HE emission is explained by the IC scattering of the BLR photons by relativistic electrons of the blob. The X-ray corona has a fixed luminosity L$_{X}$ = 0.3 $L_{d}$.  The X-ray spectrum is mainly produced by the disc-IC component for both low and high states. However, in the high state, a strong contribution from the non-thermal SSC emission from the jet is required in addition to the disc-IC emission. 

The flaring state of the source is again explained by larger Doppler factors in both scenarios. In the disc\&BLR-IC scenario the flare also features a denser blob, but this is not the case in the torus-IC solution. A 'harder when brighter' trend is also observed at $\gamma$-rays.  In both scenarios, the flaring states feature larger electron break energies.

In the torus-IC scenarios, similar to 1H\,0323$+$342, the synchrotron self-absorption is much reduced with respect to the disc\&BLR dominated scenario, due to lower electron densities. IR frequencies are described by the torus emission in both scenarios (let us recall there are not simultaneous data at these frequencies though), and the optical UV excess is well described by the multi-temperature accretion disc in both scenarios too. 

A rather large reprocessing factor of the dusty torus $\tau_{IR} \sim $ 35\% (see table \ref{pars_table_PMN}) is necessary to explain the HE emission in the torus-dominated scenario. A large corona luminosity, about half of the disc luminosity, is also required in this scenario to account for the observed X-ray emission. 

In the disc\&BLR scenario, hard X-ray emission is explained mostly by the disc IC component with minor contributions from the corona and SSC components. In the torus-IC scenario, a combination of the X-ray corona and SSC from the jet account for the hard X-ray radiation. As for 1H~0323+342, in the torus scenario the size of the emission region is in contradiction with the variability constraint we have adopted.

Because the soft X-ray excess is usually ascribed to another component whose origin is still debated (e.g. \cite{dammando_2014} and \cite{Bhattacharyya_2014}), we accept deviation of our models in this frequency band.

\begin{table*}
\caption{Model parameters for the different scenarios considered for PMN J0948+0022. The description of the model input and model-derived parameters is the same as in table \ref{pars_table_1H}.}
    \label{pars_table_PMN}
    \centering
    \begin{tabular}{lcccc}
    \hline
    Source description & \multicolumn{4}{c}{} \\
    \hline
    z & \multicolumn{4}{c}{0.5846} \\
    $\theta\, (\degree)$ & \multicolumn{4}{c}{3} \\
    $M_{BH}$ & \multicolumn{4}{c}{$1.5\times10^8$} \\
    $R_G\, [\rm{cm}]$ & \multicolumn{4}{c}{$2.22\times10^{13}$} \\
    $L_{Disc}\, [\rm{erg\, s^{-1}}]$ & \multicolumn{4}{c}{$9\times10^{45}$} \\
    $l_{Edd}$ & \multicolumn{4}{c}{0.48}  \\
    $\dot m \, [\rm{\sunmass/yr}]$ & \multicolumn{4}{c}{1.90} \\
    
    \hline
    Scenario & \multicolumn{2}{c}{Disc \& BLR} & \multicolumn{2}{c}{Torus}\\
    \cline{1-5}
    State & Average & Flare & Average & Flare \\
    
    \hline
    $\delta$ & 16 & 19 & 10 & 12 \\
    $K\, [\rm{1/cm^3}]$ & $1\times10^6$ & $4\times10^6$ & $8.2\times10^5$ & $1\times10^6$ \\
    $R_{src}\, [\rm{cm}]$ & $3.8\times10^{15}$ & $2.25\times10^{15}$ & $9.37\times10^{16}$ & $9.35\times10^{16}$ \\
    $B\, [\rm{G}]$ & 1.5 & 1.5 & 0.20 & 0.12 \\
    $n_1$ & 2.1 & 2.3 & 2.7 & 2.7 \\
    $n_2$ & 3.5 & 3.2 & 3.9  & 3.5 \\
    $\gamma_{min}$ & 10 & 10 & 450 & 450 \\
    $\gamma_{b}$ & 80 & 300 & 900 & $1\times10^3$\\
    $\gamma_{max}$ & $2\times10^4$ & $2\times10^4$ & $3\times10^4$  & $3\times10^4$ \\
    \hline
    
    $T_{IR}\, [\rm{K}]$ & 1200 & 1200 & 1200 & 1200\\
    $\tau_{IR}$ & 0.05 & 0.05 & 0.35 & 0.35 \\
    $\alpha_X$ & 1.0 & 1.0 & 1.0 & 1.0 \\
    $\tau_X$ & 0.3 & 0.3 & 0.5  & 0.5 \\
    $R_\gamma\, [\rm{R_G}]$ & $1.8\times10^3$ & $1.8\times10^3$ & $4.5\times10^4$ & $4.5\times10^4$ \\
    $\tau_{BLR}$ & 0.40 & 0.40 & $1\times10^{-4}$ & $1\times10^{-4}$ \\
    $R_{in}^{BLR}\, [\rm{R_G}]$ & $1.5\times10^4$ & $1.5\times10^4$ & $1.5\times10^4$ & $1.5\times10^4$ \\ 
    \hline
    
    $u_e/u_b$ & 17.75 & 44.96 & 5.16 &  19.80 \\
    $n_e\, [\rm{1/cm^3}]$ & $6.91\times10^4$ & $1.56\times10^5$ & 13.13 & 16.82 \\
    $u_e\, [\rm{erg\, cm^{-3}}] $ & 1.59 & 4.03 & $8.21\times10^{-3}$ & $1.13\times10^{-2}$ \\
    $L_{jet,tot}\, [\rm{erg\, s^{-1}}]$ & $4.17\times10^{46}$ & $1.13\times10^{47}$ & $8.12\times10^{45}$ & $3.15\times10^{46}$ \\
    \hline
    
    \end{tabular}
\end{table*}

\section{B2 0954+25A} \label{sect_B2}
\subsection{Observational constraints and physical properties} \label{B2_obs_constraints}
B2\,0954+25A is the most distant source of our small sample (z = 0.712, \cite{Burbidge_Strittmatter_1972}). This source is missing from general studies of the $\gamma$-ray emitting NLS1 class galaxies (e.g. \cite{Paliya_2019}, \cite{Foschini_2020_jetted_NLSy1}), certainly due to its habit of completely changing its emission spectrum. Indeed B2\,0954$+$25A has been observed in the optical in (at least) three different emission states: partially jet dominated in 1987 \citep{Jackson_Browne_1991}, disc dominated in 2004 \citep{SDSS2009}, and totally jet dominated in 2006 \citep{SDSS2006}.

Through a detailed reanalysis of the intensity and shape of the Balmer line in those spectra, \citet{Calderone2012} did find a likely value of FWHM(H{$\beta$}) $\sim$ 2800 $km s^{-1}$, just slightly above the commonly adopted classification criterion for NLS1 sources. They consequently revisited the estimation of the black-hole mass by means of virial  method and mass - stellar dispersion relation ( $M-\sigma*$), and derived a small black-hole mass, $M_{BH} = (1-3)\times10^8\sunmass$.  \citet{Calderone2012} propose this source to be a transition object between FSRQs and $\gamma$-loud NLS1s, due to its characteristics of both classes, and the strong similarity of its SED (in the disc-dominated state) to that of PMN\,J0948$+$0022.

also provides an estimate of the disc luminosity $\rm{L_{bol,Disc} = 10^{46}\ erg\ s^{-1}}$, inferred from the measurement of $H\beta$ line luminosity with a BLR  covering factor of 0.1 ($\rm{L_{bol,Disc} = 10L_{BLR}}$) and obtained a compatible value of   $1.13\times10^{46}\ \rm{erg\ s^{-1}}$ from SED modelling. The higher value of $\rm{L_{bol,Disc} = 10^{47}\ erg\ s^{-1}}$ obtained from the SDSS catalogue (\cite{Shen_2011}) is rejected due to the spectrum being taken in a jet-dominated state.

B2\,0954+25A is a compact radio-loud AGN with a flat radio spectrum. While usually flat, the radio spectrum of the source becomes inverted during burst activity (\citet{2005Torniainen}). Apparent superluminal motions (12c, \cite{Kellermann_2004}) suggest a small viewing angle $\theta$ (3 - 6\degree, \cite{Calderone2012}).

B2\,0954+25A is not as intensively studied as 1H\,0323+342 and PMN\,J0948+0022, but radio variability is also seen for this source in the MOJAVE data set at 15 GHz \footnote{\url{https://www.cv.nrao.edu/MOJAVE/}}. 

The source is known for high variability amplitudes both in optical/UV and X-ray bands (about a factor of 10 difference on timescales of $\sim$ year) as shown by the datasets presented in \cite{Calderone2012}. In the X-ray energy range, B2\,0954$+$25A has been observed by several facilities such as Einstein, \textit{Chandra}, ROSAT or \textit{Swift}-XRT, but observations are scarce and depict essentially low states, except for the Einstein 1979 observation.

Contrary to genuine $\gamma$-NLS1s which have been intensively observed, with variability features studied in depth since their discovery in $\gamma$-rays, B2\,0954+25A is poorly monitored and poorly documented, the only detailed multi-wavelength study being from  \citet{Calderone2012}. 

A monthly binned \textit{Fermi}-LAT light curve is displayed in Fig.\ref{fig:B2_multifreq_LC} showing low statistic data and a broad variability pattern. Only very few \textit{Swift} observations were taken since the launch of the \textit{Fermi} satellite. We analyzed all available \textit{Swift}-XRT X-ray observations for B2\,0954+25A together with the few simultaneous \textit{Swift}-UVOT follow-ups\footnote{see next section and Appendix for details. Data are reported in  Fig.\ref{fig:B2_multifreq_LC}}. The fractional variability is low, $F_{var} = 0.28 \pm 0.02$. However, one should keep in mind that it is the most distant source of our sample and that data only cover nine epochs, such that the observed variability might not be statistically meaningful.

\subsection{The dataset}

\textit{Chandra} observations carried out on 20 January 2009 and presented in \cite{Calderone2012} showed a low X-ray state and are used to constrain the X-ray domain of the SED as it coincides  with a low \textit{Fermi} state (Fig.\ref{fig:B2_multifreq_LC}). Because of the lack of significant variability in the $\gamma$-rays of B2 0954+25A over the years (see e.g. available FAVA\footnote{\url{https://fermi.gsfc.nasa.gov/ssc/data/access/lat/FAVA/}} or Fig.\ref{fig:B2_multifreq_LC} light curves), we consider that the long-term 4-year averaged (3FGL) \textit{Fermi} spectrum (\cite{Acero_2015}), where the source is detected with a significance of approximately 15.9$\sigma$, an integrated flux value in the 0.1 - 100 GeV energy range of $F_{1-100GeV} = (8.88 \pm 0.94)\times10^{-10}\ \rm{ph\ cm^{-2}\ s^{-1}}$, and a PL index $\Gamma = 2.44 \pm 0.07$, is a good proxy for a low state $\gamma$ spectrum. Corresponding data points are depicted in green (Fig.\ref{fig:B2_Disc_BLR} , Fig.\ref{fig:B2_Torus}). 

\textit{Fermi} did not start its operation until 2008, so no contemporary GeV data are available for the high X-ray state of 2007. Since then the source went through only two broad flaring episodes at high energy as can be seen from Fig.\ref{fig:B2_multifreq_LC}, from January-July 2010 (MJD 55200 - 55400) and in November 2016 ($\sim$ MJD 57700). Both flaring states have contemporaneous \textit{Swift} data. The flaring episode in 2016 showing a jet-dominated SED (see below), we chose to model the 2010 flare. WISE observed B2~0954+25A in  2010 May 7-8 and November 14-15, the former being contemporaneous with the flaring episode of January-July 2010 (Table \ref{tab:B20954_obs}). The average flux value of the corresponding period is plotted in orange, the second observation epoch (for which only two bands are available out of four) depicted in gray along with archival data points extracted from the ASDC database.

We analyzed the \textit{Fermi}-LAT data from the flaring epoch from January 2010 (MJD 55200) to July 2010 (MJD 55400), assuming a power-law spectral shape, a binned likelihood analysis yields a detection with a Test Statistic TS = 212 (i.e. 14.6$\sigma$), an integrated flux in the 100 MeV - 100 GeV energy range of $(5.19 \pm 0.65)\times10^{-7}\ \rm{ph\ cm^{-2}\ s^{-1}}$ and a photon index $\Gamma = 2.42 \pm 0.09$.

We also analyzed the \textit{Fermi}-LAT data from the November 2016 flaring episode. When combined with the \textit{Swift}-XRT/UVOT data, it clearly shows a jet-dominated  SED, so this state is discarded from our study.

The X-ray data used to build the SED correspond to the longest exposure ($\sim$ 7ks) observation of the source by \textit{Swift}-XRT, on the 1st of June 2007. Due to poor statistics, no good spectral information could be derived from most of the 9 epochs. The detection of this source in the X-ray band seems to require an active state. In addition to the epoch with the best statistics, we use the data points from the observation closest in time with the flaring episode, from 15 June 2010, which are depicted in red in Fig.\ref{fig:B2_Disc_BLR} and \ref{fig:B2_Torus}. Note that they both are compatible.   \textit{Swift}-UVOT simultaneous observation were taken in only one filter: UVM2($\lambda$ = 224.6nm). UVM2 band fluxes during both flaring states are almost two times larger than the two other low optical-UV states found immediately after the decay phases. 
The corresponding SED is depicted in orange (Fig.\ref{fig:B2_Disc_BLR}, Fig.\ref{fig:B2_Torus}).

The archival data used to build the SED are extracted from the ASDC database and from \cite{Calderone2012}. For completeness, all UVOT fluxes which are not simultaneous with the considered epochs are added in gray.

\begin{table}
	\centering
	\caption{Quasi-simultaneous observations of B2 0954+25A}
	\label{tab:B20954_obs}
	\begin{tabular}{lcc} 
		Instrument & Time period & MJD \\
		\hline
		\multicolumn{3}{l}{Low state} \\
		\hline
		\textit{Fermi}-LAT & 2008-2011 (3FGL) & - \\
		\textit{Chandra} & 20 Jan 2009 & 54851 \\
		\hline
		\multicolumn{3}{l}{High state} \\
		\hline
		\textit{Fermi}-LAT & Jan - Jul 2010 & 55200 - 55400 \\
		\textit{Swift} (UVOT \& XRT) & 15 Jun 2010 & 55362 \\
	    WISE & 7 - 9 May 2010 & 55323 - 55325 \\
	\end{tabular}
\end{table}

\subsection{SED modelling}

All models of the broad-band SED of B2~0954+25A assume a BH mass of $1.5\times10^8 \sunmass$, a disc luminosity $L_d = 1.13\times10^{46}\ \rm{erg\ s^{-1}}$, and a viewing angle of $\theta=3\degree$ (similar to 3.3\degree given by \cite{Homan_2021_Mojave}), as discussed in section \ref{B2_obs_constraints}. The resulting Eddington ratio, $l_{Edd} \sim$ 0.60, is relatively high, as in the case of 1H~0323+342. 

\begin{figure*}
     \centering
     \begin{subfigure}
         \centering
         \includegraphics[width=0.7\linewidth]{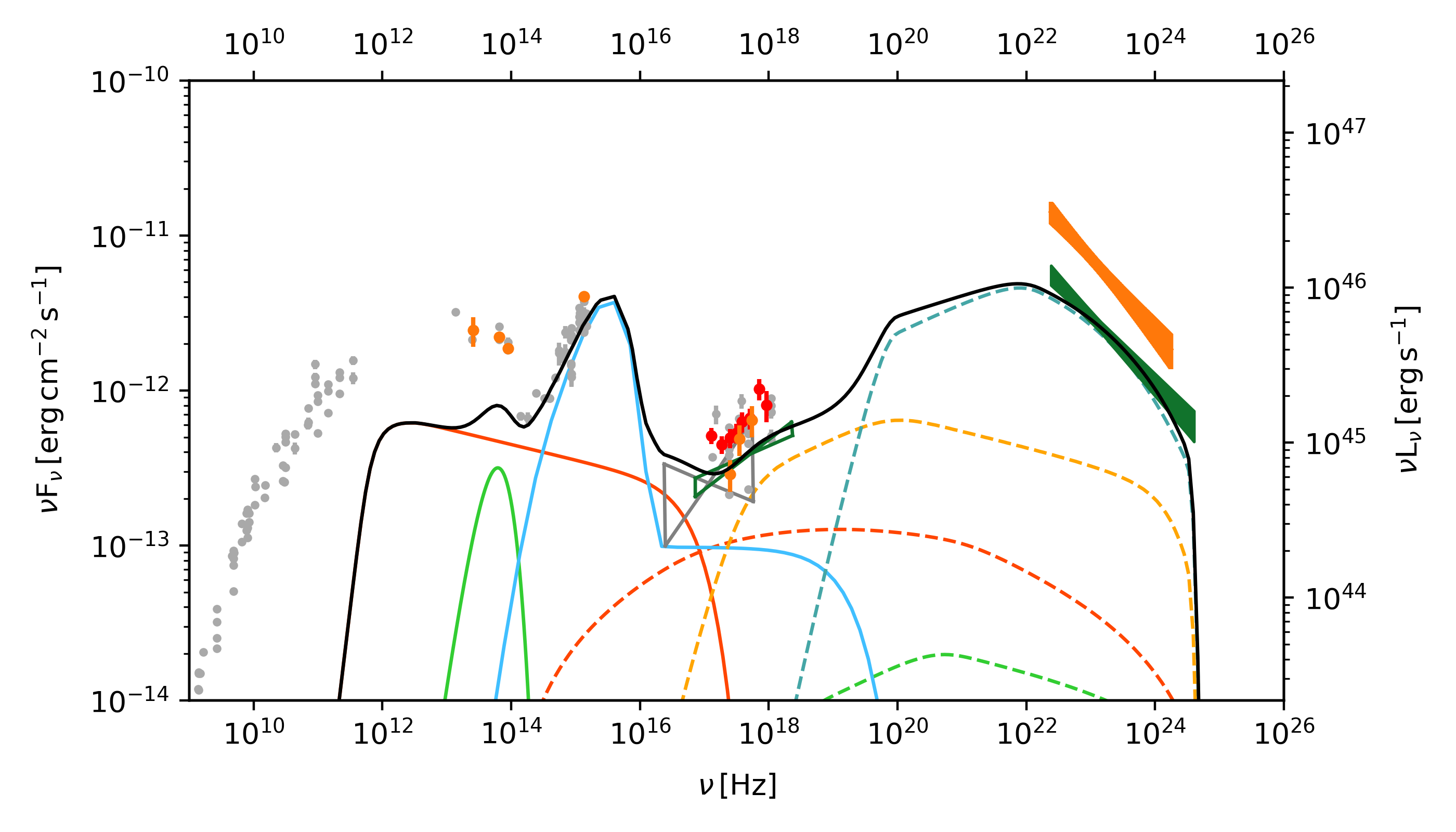}
     \end{subfigure}
     \begin{subfigure}
         \centering
         \includegraphics[width=0.7\textwidth]{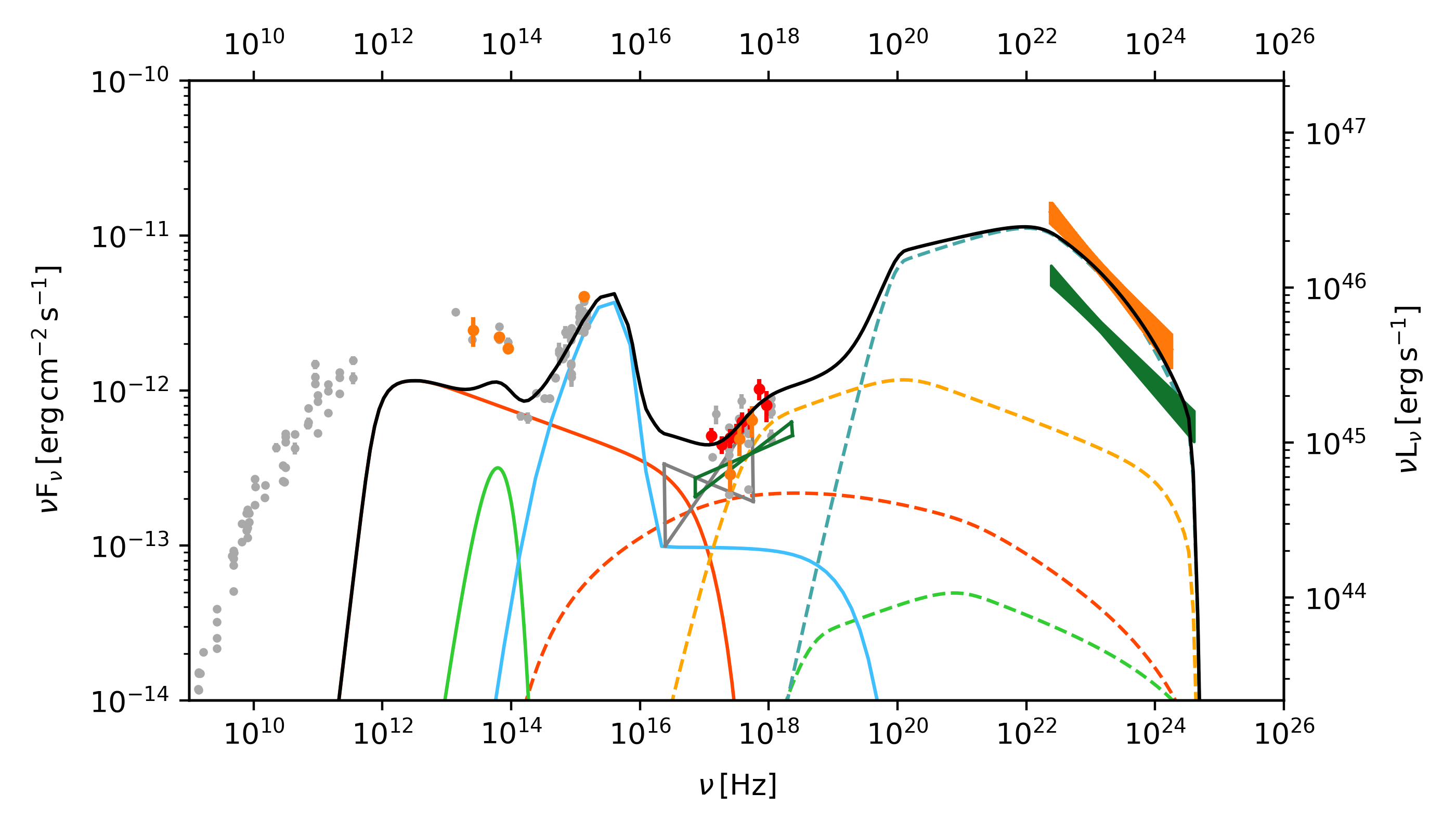}
     \end{subfigure}
        \caption{Disc\&BLR dominated scenario for B2 0954+25A. The description of the model components and the data is the same as in Fig. \ref{fig:1H_Disc_BLR}.}
        \label{fig:B2_Disc_BLR}
\end{figure*}

As for the previous sources, intermediate and the flaring states of  B2~0954+25A were modelled in a disc\&BLR-IC scenario where the blob is below the BLR and in a torus-IC scenario where the emission region is at the outermost edge of the BLR. 
The results of the broad-band modelling of the two activity states for B2~0954+25A are shown in Fig.\ref{fig:B2_Disc_BLR} and \ref{fig:B2_Torus} in the BLR- and torus-dominated scenarios, respectively and the corresponding parameters are given in table \ref{tab:pars_table_B2}.

No change of $\gamma$-ray slope being observed between the two states in the case of B2 0954+25A, the electron energy distribution remains unchanged ($n_2$ index is kept almost constant). 

In the disc\&BLR dominated case, both the average and flaring states can be reproduced by a stationary region, with a constant radius of the blob, located at a distance $R_\gamma = 2.1\times10^3\ \rm{R_G}$ from the BH. Enhanced $\gamma$-ray emission is obtained with a more relativistic and denser emitting region (higher electron number density $n_e$ and larger bulk Lorentz factor value). Slight changes in indices and break energy of the particle distribution were made for more accuracy of the fit. The  X-ray to $\gamma$-ray domain is reproduced by EIC processes (Fig.\ref{fig:B2_Disc_BLR}). IC scattering of the BLR photons by the relativistic electrons of the blob is responsible for the observed HE spectrum and the X-ray flux is mainly due to IC scattering of the disc photons. A combination of direct emission from the hot corona and SSC provides a small contribution to the soft X-ray part of the SED, compared to the dominant disc-IC emission. 

Contrary to the disc\&BLR solutions, in the torus-dominated scenario, where the HE spectrum is due to the IC scattering of the photons from the torus by the relativistic electrons of the emitting region (Fig.\ref{fig:B2_Torus}), electron Lorentz factor values become much larger compared to the disc\&BLR solutions, causing a shift in the SED towards higher energies and the non-thermal SSC emission process to account for the X-ray flux. The hot corona contributes to the observed X-ray flux in the torus-dominated case, especially in the soft X-ray band (with enhanced jet emission in the high state, Fig.\ref{fig:B2_Torus}). 

In the case of B2~0954+25A, infrared and sub-mm data are clearly underestimated by both disc\&BLR and torus-dominated solutions, as opposed to other sources for which the torus-dominated case fills the gap between the modelled synchrotron component and the archival radio-IR data, due to smaller particle densities at larger distances from the BH. It may be that B2 0954+25A requires a strong extended jet, peaking at quite high frequency,  to explain the observed radio emission, which makes B2 0954+25A very similar to Ap Librae (\cite{Hervet_2015}). 

Both BLR- and torus-dominated scenarios seem acceptable in the case of B2 0954+25A. However, the disc\&BLR solution could be preferred over the torus dominated case, given the fact that it produces a more energetic jet, in accordance with the FSRQ properties of the source (see section \ref{discussion_section_jet_powers}).

\begin{table*}
    \caption{Model parameters for the different scenarios considered for B2 0954+25A. The description of the model input and model-derived parameters is the same as in table \ref{pars_table_1H}.}
    \label{tab:pars_table_B2}
    \centering
    \begin{tabular}{lcccc}
    \hline
    Source description & \multicolumn{4}{c}{} \\
    \hline
    z & \multicolumn{4}{c}{0.712} \\
    $\theta\, (\degree)$ & \multicolumn{4}{c}{3} \\ 
    $M_{BH} [\rm{\sunmass}]$ & \multicolumn{4}{c}{$1.5\times10^{8}$} \\
    $R_G\, [\rm{cm}]$ & \multicolumn{4}{c}{$2.22\times10^{13}$} \\
    $L_{Disc}\, [\rm{erg\, s^{-1}}]$ & \multicolumn{4}{c}{$1.13\times10^{46}$} \\
    $l_{Edd}$ & \multicolumn{4}{c}{0.60} \\
    $\dot m \, [\rm{\sunmass/yr}]$ & \multicolumn{4}{c}{2.39} \\
    \hline
    Scenario & \multicolumn{2}{c}{Disc \& BLR} & \multicolumn{2}{c}{Torus}\\
    \cline{1-5}
    State & Average & Flare & Average & Flare \\
    \hline
    $\delta$ & 13 & 15 & 10 & 12 \\
    $K\, [\rm{1/cm^3}]$ & $3\times10^6$ & $5\times10^6$ & $8\times10^5$ & $2.5\times10^6$ \\
    $R_{src}\, [\rm{cm}]$ & $4.96\times10^{15}$ & $4.97\times10^{15}$ & $5.49\times10^{16}$ & $3.77\times10^{16}$ \\
    $B\, [\rm{G}]$ & 2.4 & 2.4 & 0.27 & 0.30 \\
    $n_1$ & 2.6 & 2.7 & 2.6 & 2.6 \\
    $n_2$ & 3.2 & 3.3 & 3.75 & 3.8 \\
    $\gamma_{min}$ & 10 & 10 & 500 & 500 \\
    $\gamma_{b}$ & 150 & 180 & 680 & 530 \\
    $\gamma_{max}$ & $3\times10^4$ & $3\times10^4$ & $5\times10^4$ & $5\times10^4$ \\
    \hline
    
    $T_{IR}\, [\rm{K}]$ & 1300 & 1300 & 1300 & 1300 \\
    $\tau_{IR}$ & 0.05 & 0.05 & 0.20 & 0.20 \\
    $\alpha_X$ & 1.0 & 1.0 & 1.0 & 1.0 \\
    $\tau_X$ & 0.10 & 0.10 & 0.25 & 0.25 \\
    $R_\gamma\, [\rm{R_G}]$ & $2.1\times10^3$ & $2.1\times10^3$ & $7.5\times10^4$ & $7.5\times10^4$  \\
    $\tau_{BLR}$ & 0.30 & 0.30 & $1\times10^{-4}$ & $1\times10^{-4}$ \\
    $R_{in}^{BLR}\, [\rm{R_G}]$ & $2.5\times10^4$ & $2.5\times10^4$ & $2.5\times10^4$  &  $2.5\times10^4$ \\ 
    \hline
    
    $u_e/u_b$ & 4.09 & 4.84 & 4.14 & 8.22 \\
    $n_e\, [\rm{1/cm^3}]$ & $4.77\times10^4$ & $5.96\times10^4$ & 18.10 & 46.27  \\
    $u_e\, [\rm{erg\ cm^{-3}}] $ & $9.37\times10^{-1}$ & 1.11 & $1.20\times10^{-2}$ & $2.94\times10^{-2}$ \\
    $P_{jet,tot}\, [\rm{erg\ s^{-1}}]$ & $2.06\times10^{46}$ & $3.98\times10^{46}$ & $2.41\times10^{45}$ & $4.62\times10^{45}$ \\

    \hline
    
    \end{tabular}
\end{table*}

\section{Discussion}
\label{discussion_section}

\subsection{Emission region and mechanism}
\label{discussion_section_emission}

In our scenarios, the transition from quiescent to flaring states is in general explained by a more relativistic and denser blob in the high state, with larger energy densities of the particle population and hence larger total jet powers, whereas the parameters describing the environment of the emitting region remain constant. As our solutions indicate an unchanging distance between emission region and BH for the low and flaring states, the observed variability may be explained by an enhanced plasma flow through a stationary shock region \citep{Marscher_2014}, or a shock-shock interaction between a moving and a stationary shock (\cite{Fichet2021}).

This scenario is supported by recent high-resolution VLBA observations of the $\sim (1 - 100)$ pc scale jet of 1H\,0323+342, which revealed the existence of quasi-stationary collimation and acceleration zones in the jet (\cite{Hada_2018}). Although the limited resolution of the distance scale  in that study does not allow us to strictly link their observations to our models, which assume emitting blobs in the sub-pc range, similar structures could exist in regions closer to the BH and be responsible for the acceleration of particles and observed HE emission. A similar mechanism was suggested for example for the FSRQ PKS\,1510-089 during flares in 2009 (\cite{Marscher_2009}), where a plasma knot ejected with superluminal velocity was thought to progress in an acceleration and collimation zone of the jet, then passed through a standing shock and lead to the $\gamma$-ray flare (see also \cite{Orienti_2013}).

If the parameters of the flare emission region were interpreted as a moving plasma blob, instead of enhanced emission from a stationary region, the emitting blob would be traversing the broad-line region in a few tens of days in the blob frame. In the observer frame, this would correspond to only a few days, less than the duration of the flaring states. It would be difficult to see how such a scenario could describe the observed duration of the flux variability.

For all three sources, the Lorentz factors of the particles range from $\gamma_{min} \sim 10$ to $\gamma_{max} \lesssim 10^4 $ in the disc\&BLR-dominated case and the break occurs close to the minimum value, with $\gamma_{b} \sim 10^2$. In the torus-dominated scenario, the range of electron Lorentz factors is shifted towards higher energies, with $\gamma_{min}$ values starting at 500 and slightly larger $\gamma_{max}$ too, compared to the BLR-dominated case. As a consequence, the components responsible for the X-ray emission become entirely non-thermal (SSC+corona) in the torus dominated case, compared to disc\&BLR (where a combination of disc-IC and direct emission from the corona is used to reproduce the X-ray flux). 

When estimating the Lorentz factors at which the radiative cooling break would be expected for the different scenarios and states against $\gamma_b$ applying the usual estimate \cite{Inoue1996}, one finds coherent values only for the low state of 1H\,0323+342. For all other cases, the radiative cooling break would occur well below the $\gamma_b$ values derived from modelling, implying that the electron population should already be cooled below $\gamma_b$ and that $\gamma_b$ cannot be ascribed to radiative cooling, but possibly to particle escape, to an intrinsic feature of the acceleration process, or it could reflect effects from an inhomogeneous emission region which is parameterised with a simple homogeneous model. If the particle distribution below $\gamma_b$ is already cooled, this would require a very hard particle injection spectrum with an index $\ll 2$. A possibility may be given by the re-acceleration of an already relativistic particle population, as may be expected for example in a shock-shock interaction \citep{Zech2021}. 

All our solutions are out of equipartition between electron and magnetic energy densities ${u_e} / {u_B} \sim 1$ by a factor of a few and up to an order of magnitude, as is not unusual in standard one-zone blazar emission models. To achieve equipartition in our scenarios would  require either a large number of particles or unrealistically high disc reprocessing factors due to the strong Compton dominance. High activity states are farther away from equipartition, which is not necessarily unexpected given the violent and short-term characteristic of the flares. A plasma with a low magnetisation, dominated by the particle energy density, as indicated by our scenarios, is also a requirement for efficient shock acceleration.

\subsection{Disc \& BLR vs. Torus scenario}
\label{discussion_section_scenario}

We derived the total jet luminosities from the results of the broad-band SED modelling of all sources from our sample, for both  scenarios and for all considered epochs, by computing the different contributions to the total jet power using $P_i = 2 \pi R^2 c \Gamma^2 U^{'}_{i}$. Here the primed quantity is the energy density in the co-moving frame and the index refers to the various contributions: magnetic field, radiation, relativistic electrons and cold protons\footnote{assuming one cold proton per emitting electron} and the factor 2 is for a 2-sided jet. The values are reported in Table \ref{tab:power_values_table}, showing that cold protons and radiation account for the bulk of the total jet power and that the power carried by relativistic electrons is the second most important contribution, while the magnetic flux is the least important contribution. Given the observed large Compton dominance, EIC components (EIC-BLR and EIC-Torus, in the case of BLR- and torus- dominated  scenario, respectively) account almost entirely for the total radiative power for each solution, as shown by table \ref {tab:rad_contributions}, as can be seen in the SEDs too. 

Large (external-)Compton dominance of $\gamma$-NLS1 galaxies is certainly one of the most visible similarities with FSRQs. They also share low values of the break energy $\gamma_{b}$ of the electron distribution compared to high-frequency peaked BL Lac objects.

As expected, each source and considered scenario for the EIC scattering requires much larger jet powers for flares than for quiescent states. Interestingly, the disc\&BLR dominated scenarios require significantly larger jet powers compared to the case where the torus dominates the EC radiation, except for 1H\,0323+342, for which the results are similar.

We found an increasing radiative efficiency, $\eta_{rad}$ = $P_{rad}/P_{tot}$, between the low and high states for 1H\,0323+342 and PMN\,J0948+0022, as seen by \citet{Paliya_Stalin_2016} for another $\gamma$-NLS1, PKS\,1502+036, during a GeV flare, but with a more significant difference. 
As regards B2\,0954+25A, $\eta_{rad}$ remains nearly constant between two states for a given scenario (cf. Table~\ref{tab:power_values_table}). The torus-dominated scenario for all sources and particularly for PMN\,J0948+0022 and B2\,0954+25A requires a very large radiative efficiency, between 0.5 and 0.9 for the high states, rendering this scenario less plausible. The radiative efficiency of relativistic jets is generally assumed to be around $10\%$, as is shown for example by \citet{Ghisellini_2014} for a large sample of different types of blazars. A much larger efficiency would result in jets rapidly dissipating their energy and being unable to account for the observed radio lobes and extended emission. The latter is also seen in the SEDs of the sources we consider here, so similar arguments should apply to NLS1s.

The jet emission is much lower in the X-ray band in this scenario than in the disc\&BLR one, and in turn the luminosity of the hot corona needs to be quite high, much over the 10\% usually expected. Another reason to disfavour this scenario.

Apart from energetic considerations, the torus-dominated  scenario is also problematic in terms of the size of the emission region, which exceeds by roughly an order of magnitude the limit derived from observed variability time scales for 1H\,0323+342 and PMN\,J0948+0022.

Given the degeneracy of the standard one-zone model, the values of the parameters used in the models presented above may not be unique in providing a reasonable representation of radio to $\gamma$-ray SED. However lower Doppler factors are not expected as large electron Lorentz factors are required to produce the strong EIC dominance of the SED. Radio observations showing strong core-dominated jet emission and sometimes superluminal components favor small viewing angles and large Lorentz factors too. Smaller bulk Lorentz factors would allow larger blob radii of the emitting region, located farther away from the central BH than in our cases, which would be at odds with the observed short variability timescales. Also to keep the synchrotron component below the limits and, at the same time, reproduce the powerful high-energy component, the magnetic field must be kept to relatively low values. 

\subsection{Jet powers and the nature of $\gamma-$NLS1s} \label{discussion_section_jet_powers}

\begin{figure*}
\centering
  $\vcenter{\hbox{\includegraphics[height=6cm]{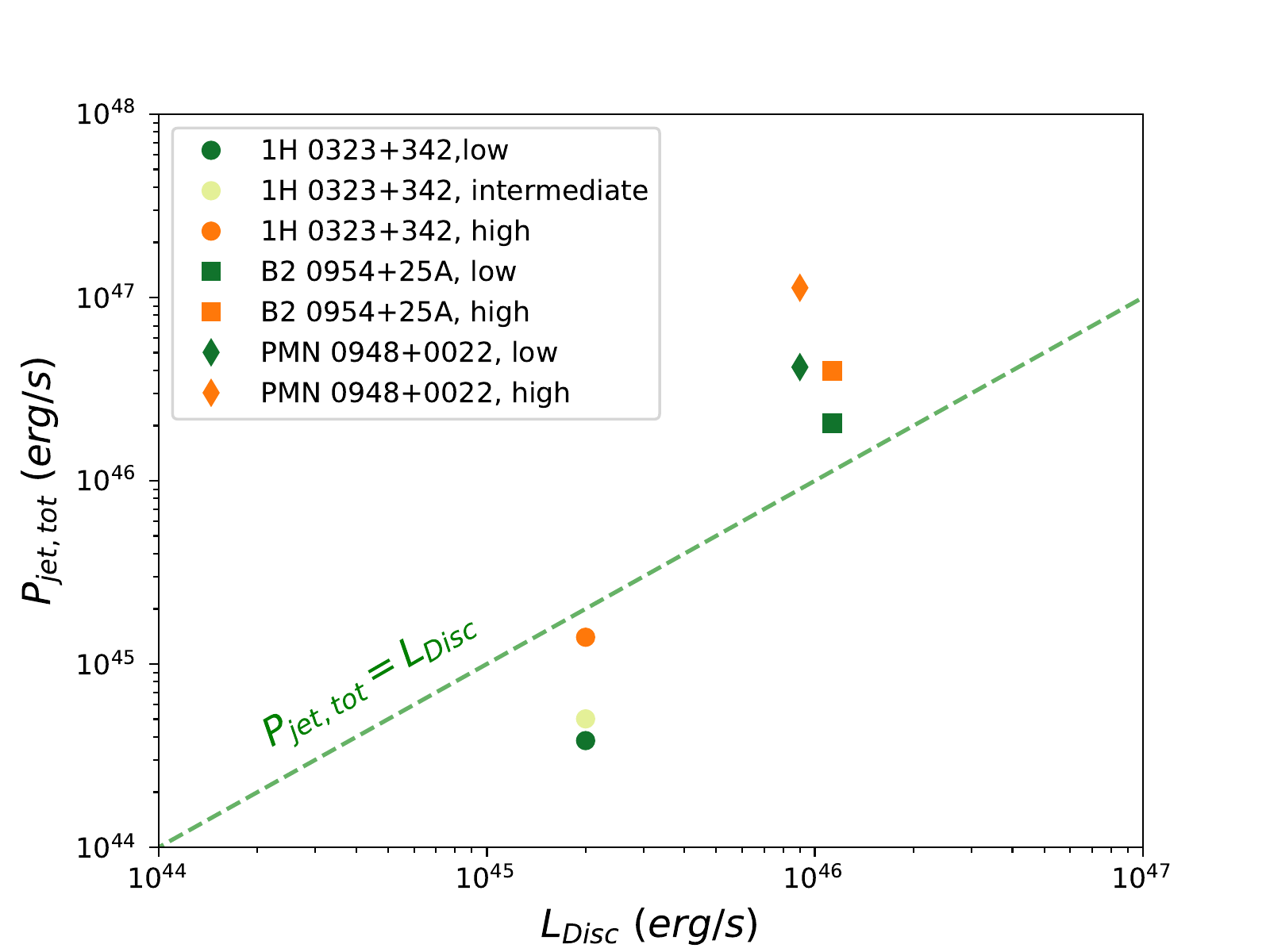}}}$
  \qquad
  $\vcenter{\hbox{\includegraphics[height=6cm]{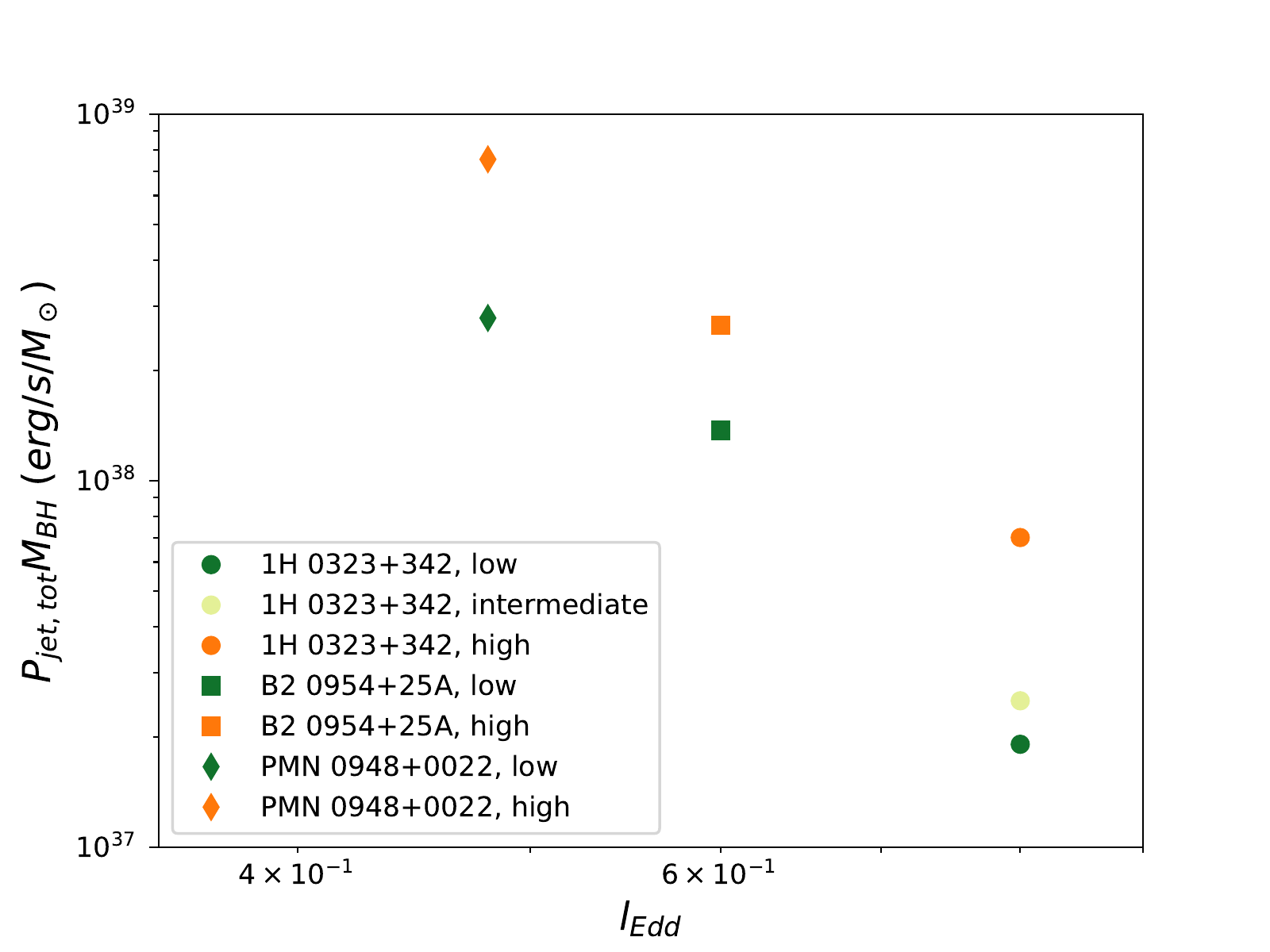}}}$
\caption{Total jet power vs disc luminosity (left panel) and total jet power corrected for BH mass vs Eddington ratio (right panel) for the disc \& BLR dominated scenario.}
\label{fig:characterisation_upper_panel}
\end{figure*}

The left panel of Fig.\ref{fig:characterisation_upper_panel} represents the total jet powers estimated from SED modelling of different sources and states, as a function of the disc luminosity, for the disc\& BLR-dominated scenario (see Fig. \ref{fig:characterisation_upper_panel_torus} for the torus-dominated scenario).

The increase of total jet powers with accretion luminosities is expected, if one assumes that the disc feeds the jet or that both depend on the black hole mass and spin. 
Our preferred solutions for B2\,0954+25A and PMN\,J0948+0022 have jet powers that are up to an order of magnitude larger than the disc luminosity, as was also the case for PKS\,J1222$+$0413, a FSRQ tentatively reclassified as a NLSy1 (\cite{Kynoch_2019}). Strong dominance of the jet power over the disc luminosity, as seen here, was reported by \citet{Ghisellini_2014} for different types of blazars. On the other hand, \citet{Paliya_2016} showed that radio-loud NLS1s, including some $\gamma$-ray emitters, are characterized by a different trend in the $P_{jet,tot}$ vs $L_{Disc}$ diagram and that their jet powers are in most cases below the disc luminosity, indicative of higher accretion rates, radiatively efficient discs and relatively low power jets.  

In this respect, 1H\,0323+342, with a jet power below the disc luminosity, is more typical
of the NLS1 population than the other two sources. 1H\,0323+342 has the least powerful jet and harbors the least massive BH of the three sources (one order of magnitude below those of B2\,0954+25A and PMN\,J0948+0022).

The results obtained from the present multi-epoch study of the SEDs suggest that 1H\,0323+342 remains moderately powerful, regardless of the activity state and adopted scenario. 

We find that the jet powers derived from the SED modelling are systematically larger (by almost an order of magnitude) than the estimates provided by \cite{Foschini_2015}, deduced from the radio core measurements at 15 GHz (see \cite{Foschini_2014}) (log($P_{jet,tot}$) = 43.60 for 1H~0323+342 and 45.38 for PMN~J0948+0022). The statistical relations used to estimate the jet powers from the radio data are based on a unified view of the astrophysical jets, inducing large associated uncertainties, while  jet power estimates from SED modelling are obtained separately for each particular source and epoch.

For 1H\,0323+342, when comparing the jet power $log(P_{jet})$ = 44.58 of the low state in our Disc and BLR-dominated scenario to the same {\it Fermi}-LAT state modelled by \cite{Kynoch_2017}, we find a result that is larger by less than a factor of 3 with $log(P_{jet})$ = 45.01. To model the different activity states with a single zone, the photon field from the accretion disc is not sufficient, so we have to place the emission region farther away from disc than \cite{Kynoch_2017}, leading to a larger jet power. \cite{Zhang_2020} have also published jet power estimates from radiative models for a range of activity states of the source, yielding a similar range of values ($8 \times 10^{43}$ to $3 \times 10^{44}\ \rm{erg\ s^{-1}}$) to ours ($8 \times 10^{43}$ to $5 \times 10^{44}\ \rm{erg\ s^{-1}}$) for the jet power excluding the hadronic contribution. While the overall jet powers are very similar, their solutions present a higher magnetic contribution than ours. 
\cite{D'Ammando_2015} do not provide $P_{jet}$ for their solution, but we can compare the values of the individual components. Our $P_e$ in the BLR-dominated scenario ($\sim 4\times10^{44}\ \rm{erg\ s^{-1}}$) is similar to their value ($5.48\times10^{44}\ \rm{erg\ s^{-1}}$), while $P_B$ and $P_{p,cold}$ are respectively smaller and larger by an order of magnitude. 

Comparing our preferred disc\&BLR-dominated scenario to the result from \cite{Paliya_2019} for the intermediate state of PMN\,J0948+0022, our jet power is smaller by a factor of 3 ($log(P_{jet})$ = 46.62 vs. 47.11), with a considerably smaller contribution from the magnetic field. In our model, the location of the emission region is closer to the central black hole and the disc luminosity is higher, which should result in a higher contribution from the EC-disc component and reduce the jet power
requirement. An estimate for the jet power for the high state of this source is given again by \cite{Zhang_2020}, at about a factor 10 smaller than our result. This large difference seems mainly due to a very different shape of the high-energy emission bump, which is much broader in our case.

For B2\,0954+25A, only an estimate for a state similar to our low state was found in the literature \citep{Calderone2012}, yielding a very similar jet power to our result, both about $1.5 \times 10^{45}$ erg/s, but again with a significantly larger magnetic contribution than in our scenario. This is due to a higher magnetic field in the emission region of the model by \cite{Calderone2012}, which assumes also a larger distance from the central black hole.

Given the large number of free parameters in multi-component EIC models, it is of course difficult to draw general conclusions from these comparisons. The limited observational constraints on the synchrotron emission from the compact blob and the different possible combinations of the EIC contributions to the high-energy component leave, for example, a large range of possible locations of the emission region or strength of the magnetic field.

To verify among our sources whether $\gamma$-NLS1 host intrinsically lower power jets or are rather low-mass analogs of blazars, we normalized the jet powers by the black-hole masses (cf. right panel of Fig.\ref{fig:characterisation_upper_panel} and \ref{fig:characterisation_upper_panel_torus}). 
In the disc\&BLR-dominated scenario, 1H\,0323+342 has the largest Eddington ratio and lowest  jet power, even after scaling it by the BH mass, for both low and high states, which supports the conclusions by \cite{Kynoch_2017} about its intrinsically low power jet. The other two sources have comparable jet powers after scaling, both at 
an order of magnitude higher level than 1H\,0323+342.
These conclusions would not hold for the disfavoured torus-dominated scenario.

\begin{table*}
    \centering
    \begin{tabular}{l|c|c|c|c|c|c|c|c}
          Source & Scenario & State & log$P_e$ &  log$P_B$ & log$P_{rad}$ & log$P_{p,cold}$ &  log$P_{tot,jet}$ & $\eta_{rad}$ \\
                     \hline
                     \hline
         \multirow{6}{*}{1H 0323+342} & \multirow{3}{*}{Disc-BLR} & Low  & 43.20 & 42.21 & 43.80 & 44.48 & 44.58 & 0.17 \\
                                      &                           & Intermediate & 43.32 & 42.21 & 43.97 & 44.58 & 44.70 & 0.19 \\         
                                      &                           & High & 43.78 & 42.31 & 44.65 & 44.94 & 45.15 & 0.32 \\
                                      & \multirow{3}{*}{Torus}    & Low  & 43.66 & 42.62 & 43.85 & 44.08 & 44.38 & 0.30 \\
                                      &                           & Intermediate & 43.62 & 42.63 & 43.91 & 44.00 & 44.36 & 0.36 \\
                                      &                           & High & 44.38 & 42.95 & 44.91 & 44.75 & 45.22 & 0.50 \\
         \hline
        
         \multirow{4}{*}{PMN J0948+0022} & \multirow{2}{*}{Disc-BLR} & Intermediate  & 44.67 & 43.42 & 46.04 & 46.48 & 46.62 & 0.26 \\
                                      &                              & High & 45.08 & 43.43 & 46.62 & 46.85 & 47.05 & 0.37 \\
                                      & \multirow{2}{*}{Torus}       & Intermediate  & 44.60 & 43.89 & 45.82 & 44.99 & 45.91 & 0.82 \\
                                      &                              & High & 44.93 & 43.64 & 46.46 & 45.28 & 46.50 & 0.91 \\
     
        \hline
        \multirow{4}{*}{B2 0954+25A} & \multirow{2}{*}{Disc-BLR} & Low  & 44.39 & 43.78 & 45.14 & 46.28 & 46.31 & 0.07 \\
                                      &                           & High & 44.65 & 43.96 & 45.52 & 46.56 & 46.60 & 0.08 \\
                                      & \multirow{2}{*}{Torus}    & Low  & 44.30 & 43.69 & 45.23 & 44.66 & 45.38 & 0.71 \\
                                      &                           & High & 44.56 & 43.64 & 45.53 & 44.93 & 45.67 & 0.73 \\

        \hline
        \hline
    \end{tabular}
    \caption{Physical quantities derived, assuming a 2 sided jet, from the modelling of the sources. All quantities are given in units of $erg/s$. Different contributions to the total jet power $P_{tot,jet}$: Poynting flux power ($P_B$), total radiative power $P_{rad}$, the power carried out by relativistic electrons ($P_e$) and cold protons $P_{p,cold}$. $\eta_{rad}$ defined as $P_{rad}/P_{jet,tot}$ refers to the radiative efficiency.}
    \label{tab:power_values_table}
\end{table*}

\cite{Kynoch_2017} demonstrated that 1H\,0323+342 cannot be a 'mini' version of FSRQs since the scaling of the FSRQ model to lower BH masses yields larger jet power than the one derived, despite the similarities of its SED properties with the FSRQ class. Similar conclusions were drawn by \cite{Abdo_2009_Discovery} and \cite{Paliya_2014} when comparing with PMN\,J0948+0022, which in our preferred solution does not differ significantly from the intermediate source B2\,0954+25A in terms of scaled jet power.

At the same time, the violent change of the jet powers between low and high states, which is visible from Fig.~\ref{fig:characterisation_upper_panel} particularly for 1H\,0323+342 and PMN\,J0948+0022, reflects that these sources are capable of transiting from a low power regime to a blazar-like phase, accounting for their variable spectral features.

The double Seyfert-blazar nature of $\gamma$-NLS1s is also supported by their X-ray properties. $\gamma$-NLS1s have different X-ray signatures from NLS1 galaxies and are more similar to FSRQs, since they are dominated by the emission from the  jet in this frequency range (e.g. \cite{Paliya_2019}, \cite{D'Ammando_2020_Swift_view}). The results of the modelling of our low states show that X-ray data are mainly reproduced by the direct emission from the corona, as also observed for NLS1 galaxies, whereas in the high state non-thermal emission from the jet becomes predominant.

In addition, \textit{Fermi} analysis of $\gamma$-NLS1 galaxies performed by \cite{Paliya_2015} reported steeper indices of the HE spectra for these sources, compared to generally harder spectra of blazars (see \cite{Paliya_2019}), which was interpreted as one of the reasons explaining only few detections of $\gamma$-NLS1 in HE. However, the flaring states modelled in the present study show a perhaps more blazar-like behavior with quite hard HE spectra for PMN\,J0948+0022 and 1H\,0323+342, which is suppressed by the analysis of long-term averaged states. 

$\gamma$-NLS1 galaxies have not been detected in the VHE band by the existing Cherenkov telescopes. \citet{D'Ammando_2015} reported no significant $\gamma$-ray signal from PMN\,J0948+0022 observations by VERITAS  following the most powerful flare of PMN\,J0948+0022 detected by {\it Fermi}-LAT at the end of 2012. Extrapolation to the VHE band of the LAT measured energy spectrum averaged over the active phase of the flare indeed falls below the VERITAS sensitivity in 5 h of observation. The probability of the detection by the next generation of Cherenkov telescopes, the upcoming CTA project, of the most promising sources of this class was investigated by \cite{Romano_2020}. According to their results, the vast majority of known $\gamma$-NLS1 would remain undetected, even with the sensitivity of the CTA.  PMN\,J0948+0022 , if in a flaring state and within some conditions (source located below the BLR, for a specific model of the BLR, see the paper for more details) is one of the 3 that could be detectable.

\section{Conclusions and Outlook}

We have investigated the main physical properties of two $\gamma$-NLS1 galaxies and one intermediate object between FSRQ and  $\gamma$-NLS1 through the broad-band SED modelling of different activity states, in order to investigate the origin of their $\gamma$-ray flares. We analysed their spectral properties, and assessed the energetics of their jets, at the different considered epochs. \\

Our main conclusions are as follows:

\begin{itemize}
\item The overall SEDs of the three sources can be explained with a compact 
 stationary emission region inside the jet, a contribution from the BLR and dust torus, a standard optically thick, geometrically thin emission disc and a corona. The dominant high-energy
 emission is explained by the EIC processes dominated either by photon fields from the disc and BLR or from the torus, depending on the assumed distance of the emission region.
 An additional emission from the extended jet would be needed to fully account for the radio emission.

\item The direct emission from the corona contributes to the X-ray mostly during low/intermediate states in $\gamma$-NLS1 galaxies, whereas during flaring episodes, the jet emission contributes the most in this frequency band.
This leads to a changing character of the SED - resembling more the emission
from Seyfert-1s in the low states and that of blazars in the high states.

 \item The scenarios in which EIC emission is dominated by photon fields from the disc and BLR are preferred for all three sources due to the large radiative efficiency ($\gg 10$ \%) for the jet emission that would be required in the torus dominated scenarios. 

\item The transition between low, intermediate and high activity states in $\gamma$-rays is explained by changes in the jet emission region only, with a denser, and more relativistic ``blob`` with a harder electron spectrum in the high states. The parameters describing the surrounding environment of the blob are kept mostly unchanged. 

\item A physical scenario to account for the flaring states with a stationary
source and the energy distribution resulting from our models could be given 
by the interaction of moving and a standing shock in which relativistic particles are re-accelerated. 

\item While the deduced jet power for the preferred disc\&BLR-dominated scenario is comparable between the NLS1 PMN\,J0948+0022 and the FSRQ/intermediate object
B2\,0954+25A, the other NLS1 1H\,0323+342 exhibits a significantly lower jet power, even when scaled to the black hole mass.

\end{itemize}

\section*{Acknowledgements}

Part of this work is based on archival data, software or online services provided by the ASI Space Science Data Center, the Astrophysics Science Archive Research Center (HEASARC), the Astrophysics Data Systems (ADS). This research has made use of the SIMBAD database, operated at CDS, Strasbourg, France. This research has also made use of the NASA/IPAC Infrared Science Archive, which is funded by the National Aeronautics and Space Administration and operated by the California Institute of Technology (doi: 10.26131/IRSA1 - All WISE Source Catalog).

\section*{Data Availability}
The data analysis results presented in this paper are available upon request to the authors.


\bibliographystyle{mnras}
\bibliography{example} 




\appendix

\section{Multi-frequency light curves)}
\textit{Fermi} data are retrieved from \textit{Fermi} light curve repository. \footnote{\url{https://fermi.gsfc.nasa.gov/ssc/data/access/lat/LightCurveRepository/}}

\begin{figure*}
    \centering
    \includegraphics[width=1.\linewidth]{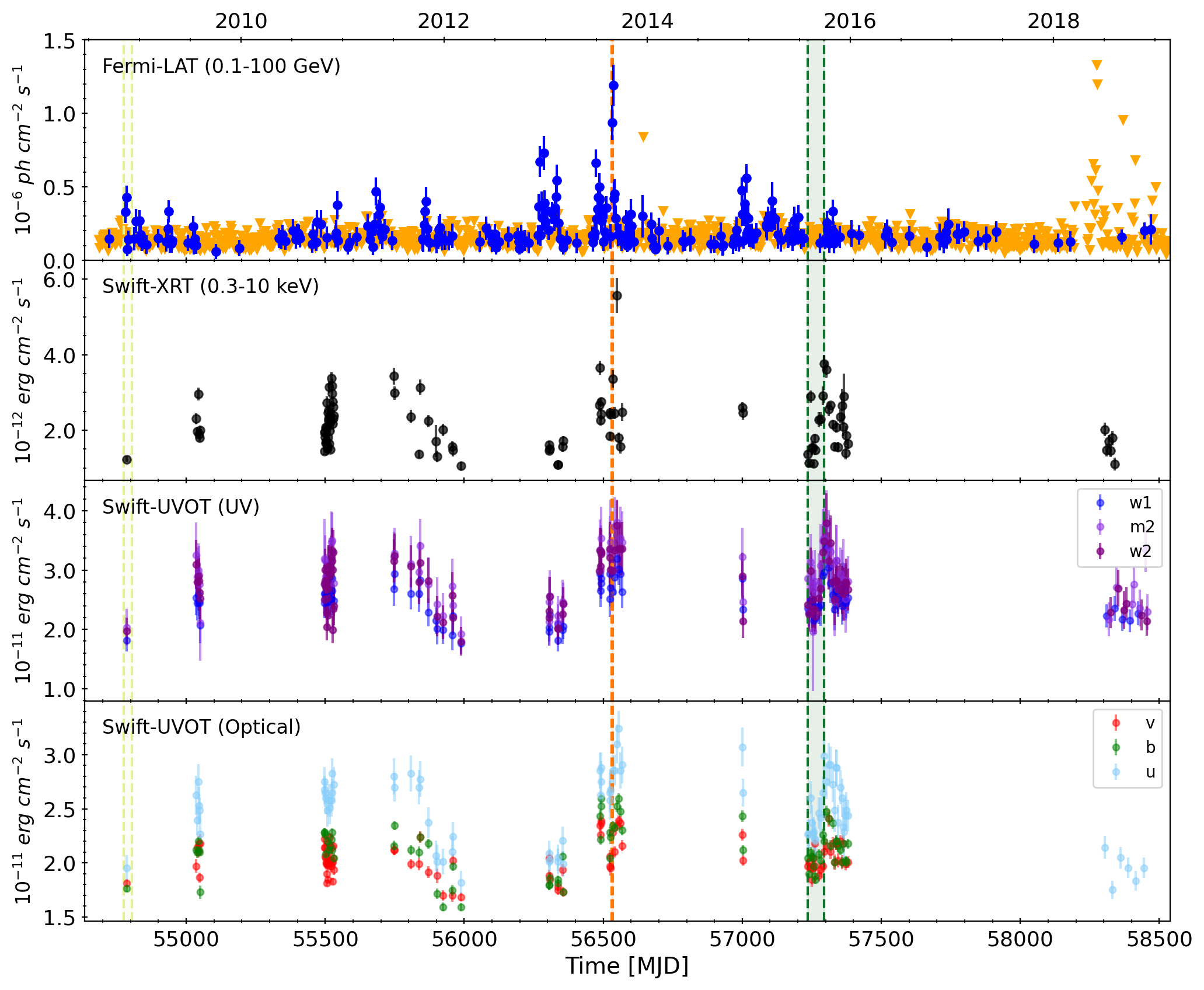}
    \caption{Multi-frequency light curves of 1H 0323+342. Upper panel: 3 day binned \textit{Fermi}-LAT light curve, photon flux in the 0.1 - 100 GeV energy range, in units of $10^{-6}\ \rm{ph\ cm^{-2}\ s^{-1}}$. Middle panel: XRT flux in the $0.3 - 10\ \rm{keV}$ energy range, in units of $10^{-11}\ \rm{erg\ cm^{-2}\ s^{-1}}$. Lower panel: UVOT flux in the different optical and UV bands of \textit{Swift}, in units of mJy. \textit{Swift} XRT and UVOT data are retrieved from \citet{D'Ammando_2020_Swift_view}. For more visibility, the entire light curves are not shown and they are truncated to focus on the periods of interest.}
    \label{fig:1H_multifreq_LC}
\end{figure*}

\begin{figure*}
    \centering
    \includegraphics[width=1.\linewidth]{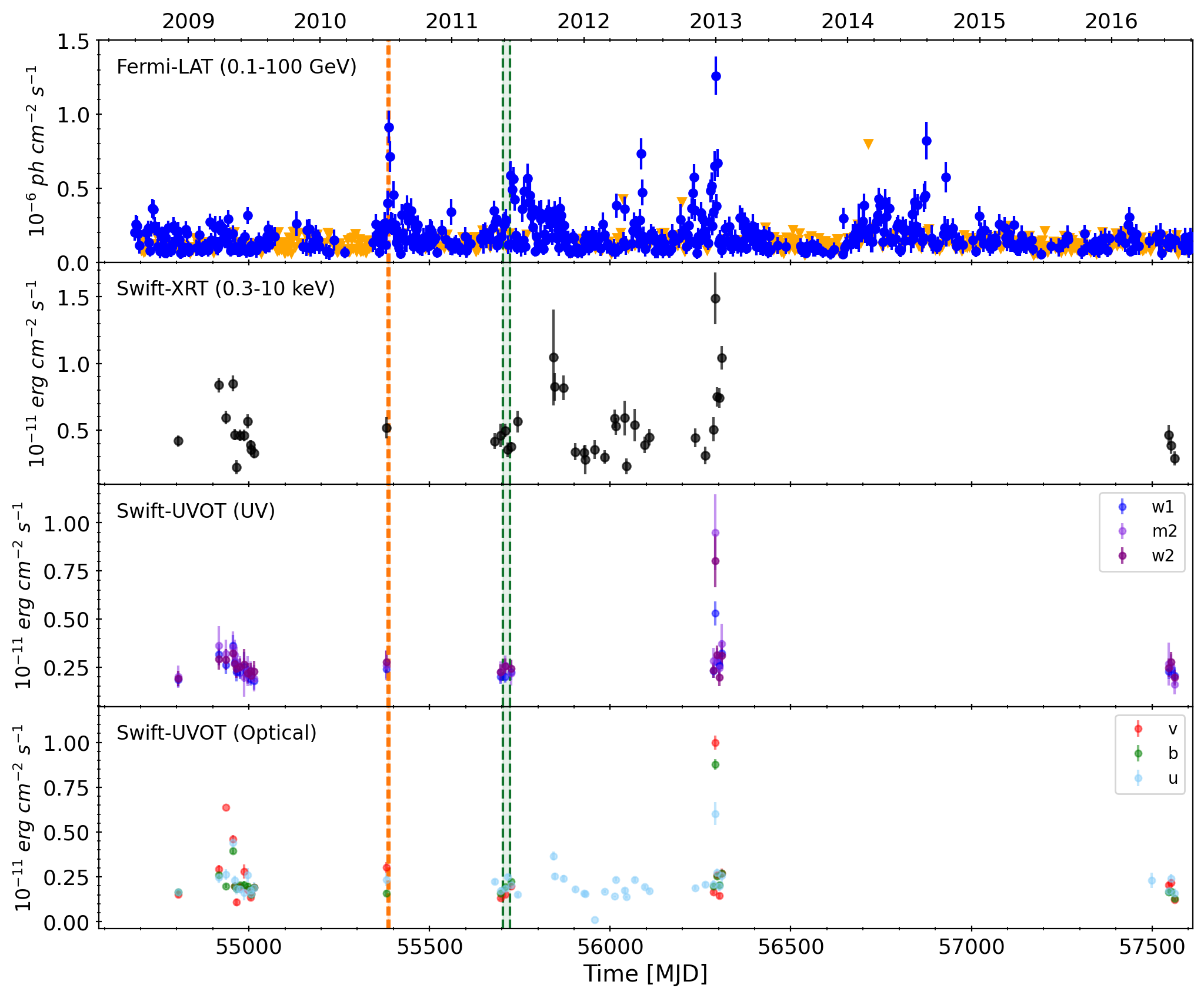}
    \caption{Multi-frequency light curves of PMN J0948+0022. Upper panel: 3 day binned \textit{Fermi}-LAT light curve, photon flux in the 0.1 - 100 GeV energy range, in units of $10^{-6}\ \rm{ph\ cm^{-2}\ s^{-1}}$. Middle panel: XRT flux in the $0.3 - 10\ \rm{keV}$ energy range, in units of $10^{-12}\ \rm{erg\ cm^{-2}\ s^{-1}}$ Lower panel: UVOT flux in the different optical and UV bands of \textit{Swift}, in units of mJy. \textit{Swift} XRT and UVOT data are retrieved from \citet{D'Ammando_2020_Swift_view}. For more visibility, the entire light curves are not shown and they are truncated to focus on the periods of interest.}
    
    \label{fig:PMN_multifreq_LC}
\end{figure*}

\begin{figure*}
    \centering
    \includegraphics[width=1.\linewidth]{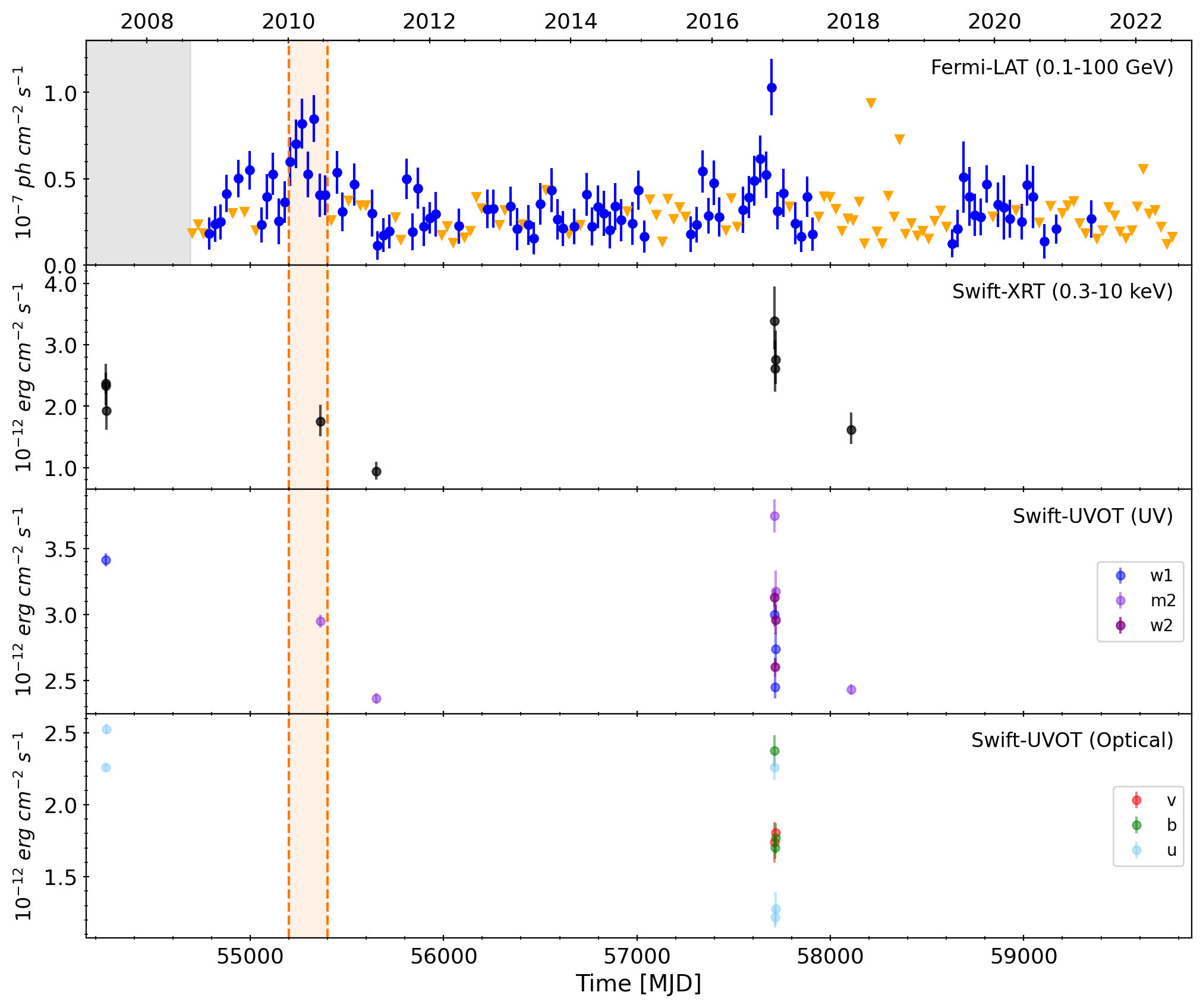}
    \caption{Multi-frequency light curves of B2 0954+25A. Upper panel: monthly-binned \textit{Fermi}-LAT light curve, photon flux in the 0.1 - 100 GeV energy range, in units of $10^{-7}\ \rm{ph\ cm^{-2}\ s^{-1}}$. Middle panel: XRT flux in the $0.3 - 10\ \rm{keV}$ energy range, in units of $10^{-12}\ \rm{erg\ cm^{-2}\ s^{-1}}$. Lower panel: UVOT flux in the different optical and UV bands of \textit{Swift}, in units of mJy.  The grayed area delimits the period before the start of \textit{Fermi} observations. Note that we only show the flaring state, since the modelled low state refers to the entire 3FGL flux.}
    \label{fig:B2_multifreq_LC}
\end{figure*}

\section{Data analysis} \label{sect_data_analysis}

\subsection{\textit{Fermi}-LAT data}

We downloaded LAT data from the \textit{Fermi} Science Support Center archive\footnote{\url{https://fermi.gsfc.nasa.gov/cgi-bin/ssc/LAT/LATDataQuery.cgi}}  for the different targets and time periods of interest and considered all photons in a circular region of interest (RoI) of radius 10°, centred on the source position. A zenith-angle cut of 90° was applied. The analysis was done with the \textit{Fermi}  ScienceTools software package\footnote{\url{https://fermi.gsfc.nasa.gov/ssc/data/analysis/software/}}  version v11r5p3, in combination with PASS 8 instrument response functions (event class 128 and event type 3) corresponding to the P8R3\_SOURCE\_V3 response and the gll\_iem\_v07.fits and iso\_P8R3\_SOURCE\_V3\_v1.txt models for the Galactic and diffuse background respectively. The source model used include all sources of the 4FGL catalogue (\cite{4FGL_paper}) that fall within 15° of the RoI center. A binned likelihood analysis was applied in an iterative way, fixing the spectral shape parameters for 4FGL sources more than 10° away  to account for event leakage in the RoI due to the large PSF at low energies. In a second step, the sources contributing to less than 5 per cent of the total number of counts in the RoI and/or  with significance, derived from test statistic, $TS < 9$ have their parameters frozen. In the end the only free parameters  are those of sources less than 3° away from the target, if not frozen in the previous step and the normalizations of the two diffuse background components.  

All the uncertainties presented in this paper are statistical only.

\subsection{\textit{Swift} analysis - B2~0945+25A}
The Neil Gehrels \textit{Swift} Observatory (\citet{Gehrels_XRT_2004}) performed only nine observations of B2~0945+25A in 18 years (see table \ref{tab:XRT_B2}).  The hard X-ray flux of this source is below the sensitivity of the BAT instrument for the short exposures of these observations, also the source is not present in the Swift BAT 105-month hard X-ray catalogue (\citet{Oh_BAT_2018}). We thus only analyzed all the X-ray Telescope (XRT, 0.2–10.0 keV) and the Ultraviolet/Optical Telescope (UVOT) available data. 

The XRT telescope data were processed with the XRTDAS software package  within the HEASoft 6.29c package\footnote{\url{https://heasarc.gsfc.nasa.gov/docs/software/lheasoft/download.html}}. Event files were calibrated and cleaned with the default filtering criteria with the xrtpipeline v0.13.6 task.  
The spectrum was extracted from the summed and cleaned event file with the xselect task. The source extraction region was a 47.2 arcsec radius centred on the source,  while background events were extracted from a circular region with radius of 70 arcsec away from the source region and any other source. The ancillary response files  were generated with the xrtmkarf task and the latest response matrices available in the \textit{Swift} CALDB were used. As the spectra have a low numbers of photons, before the spectral fitting, the source spectra were grouped with the grppha task to ensure a minimum of 1 count per bin and the fit performed with the Cash statistic (\citet{Cash_1979}). 

The spectrum of each \textit{Swift}-XRT observation was fitted with a simple power law with the hydrogen column fixed to the  Galactic absorption column of $3.47 \times 10^{20}$ atoms cm$^{-2}$ using the XSPEC 12.9.1 package.  The resulting photon index and de-absorbed flux in the 0.3–10 keV energy are given in table \ref{tab:XRT_B2}. 

During the \textit{Swift} pointings, the UVOT instrument observed the source in at least one of its 6  photometric bands\footnote{V (500–600 nm), B (380–500 nm), U (300–400 nm), UVW1 (220–400 nm), UVM2 (200–280 nm) and UVW2 (180–260 nm)}.  An aperture photometry analysis was performed. Separate images within a given file were integrated with the uvotimsum task and then analyzed by using the uvotsource task. Source counts were extracted from a circular region of 5 arcsec radius centred on the source, while background counts were derived from a 20 arcsec radius region located in a nearby source-free area. The UVOT magnitudes are corrected for Galactic extinction using $E(B-V) = 0.0325$\,mag and the extinction laws from \citet{Cardelli89} and converted to flux densities using the conversion factors from \citet{Breeveld10}.

\begin{table*}
    \caption{Summary of the \textit{Swift}-XRT observations of B2 0954+25A.}
    \label{tab:XRT_B2}
    \begin{tabular}{ccccccc}
         \hline
         Date       & MJD start & Obs. ID     &Exposure & Index & Flux(0.3-10 keV)       \\
            &           &             & ks      &       &  10$^{-12}$ erg cm$^{-2}$ s$^{-1}$     \\
         \hline                                                                          
         2007-05-31 & 54251.69  & 00036325001 & 2.0     &  1.69$\pm$0.16   & 2.33$\pm$0.36    \\
         2007-06-01 & 54252.03  & 00036325002 & 7.6     &  1.73$\pm$0.08   & 2.37$\pm$0.18    \\
         2007-06-05 & 54256.05  & 00036325003 & 2.0     &  1.62$\pm$0.20   & 1.93$\pm$0.36    \\
         2010-06-15 & 55362.21  & 00036325004 & 2.6     &  1.76$\pm$0.18   & 1.75$\pm$0.27    \\
         2011-03-30 & 55650.26  & 00036325005 & 4.8     &  1.82$\pm$0.20   & 0.94$\pm$0.16    \\
         2016-11-18 & 57710.88  & 00036325006 & 2.0     &  1.46$\pm$0.16   & 3.39$\pm$0.55    \\
         2016-11-22 & 57714.47  & 00036325007 & 2.0     &  1.52$\pm$0.18   & 2.62$\pm$0.45    \\
         2016-11-24 & 57716.20  & 00036325008 & 1.9     &  1.63$\pm$0.18   & 2.76$\pm$0.48    \\
         2017-12-19 & 58106.83  & 00088202001 & 2.3     &  1.80$\pm$0.20   & 1.62$\pm$0.28    \\ 
         \hline         
    \end{tabular}
\end{table*}

\begin{table*}
    \caption{Summary of the \textit{Swift}-UVOT observations of B2 0954+25A.}
    \begin{tabular}{cccccccccc}
        \hline
         Date  & MJD start & Obs. ID & $n_{images}$ & U & B & V & UVW1 & UVM2 & UVW2 \\
        \hline
         2007-05-31 & 54251.69  & 00036325001 & 5 &     &      &       & 16.53$\pm0.04$       &      &              \\
         2007-06-01 & 54252.03  & 00036325002 & 10&17.08$\pm$0.04    &      &       &        &      &              \\
         2007-06-05 & 54256.05  & 00036325003 & 3 &17.01$\pm0.04$     &      &       &        &      &              \\
         2010-06-15 & 55362.21  & 00036325004 & 1 &     &      &       &        &17.72$\pm$0.04      &              \\
         2011-03-30 & 55650.26  & 00036325005 & 1 &     &      &       &        &16.96$\pm$0.04      &              \\
         2016-11-18 & 57710.88  & 00036325006 & 1 &17.17$\pm$0.07     &18.06$\pm$0.09      &17.56$\pm$0.09       &16.94$\pm$0.07        &16.61$\pm$0.07      &16.95$\pm$0.05             \\
         2016-11-22 & 57714.47  & 00036325007 & 1 &17.61$\pm$0.06     &18.27$\pm$0.07      &       &17.02$\pm$0.06       &      &17.14$\pm$0.06              \\
         2016-11-24 & 57716.20  & 00036325008 & 2 &17.57$\pm$0.08     &18.19$\pm$0.09      &18.17$\pm$0.20       &17.05$\pm$0.07        &16.75$\pm$0.07      &17.03$\pm$0.06              \\
         2017-12-19 & 58106.83  & 00088202001 & 3 &     &      &       &        & 16.96$\pm$0.05    &              \\ 
        \hline
    \end{tabular}
\end{table*}


\onecolumn
\section{Torus-EIC model results} \label{sect_torus_models}

\begin{figure*}
     \centering
     \begin{subfigure}
         \centering
         \includegraphics[width=0.7\linewidth]{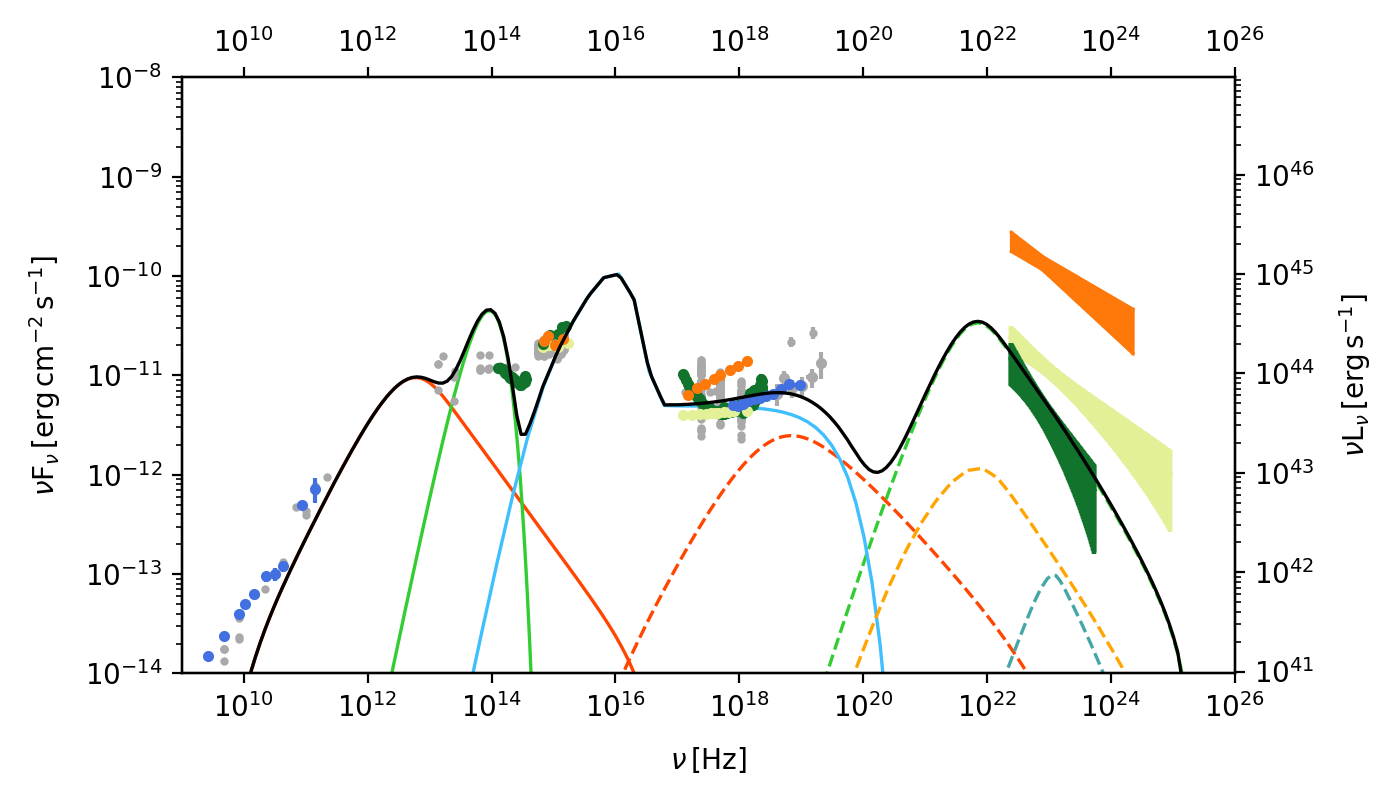}
     \end{subfigure}
     
     \begin{subfigure}
         \centering
         \includegraphics[width=0.7\linewidth]{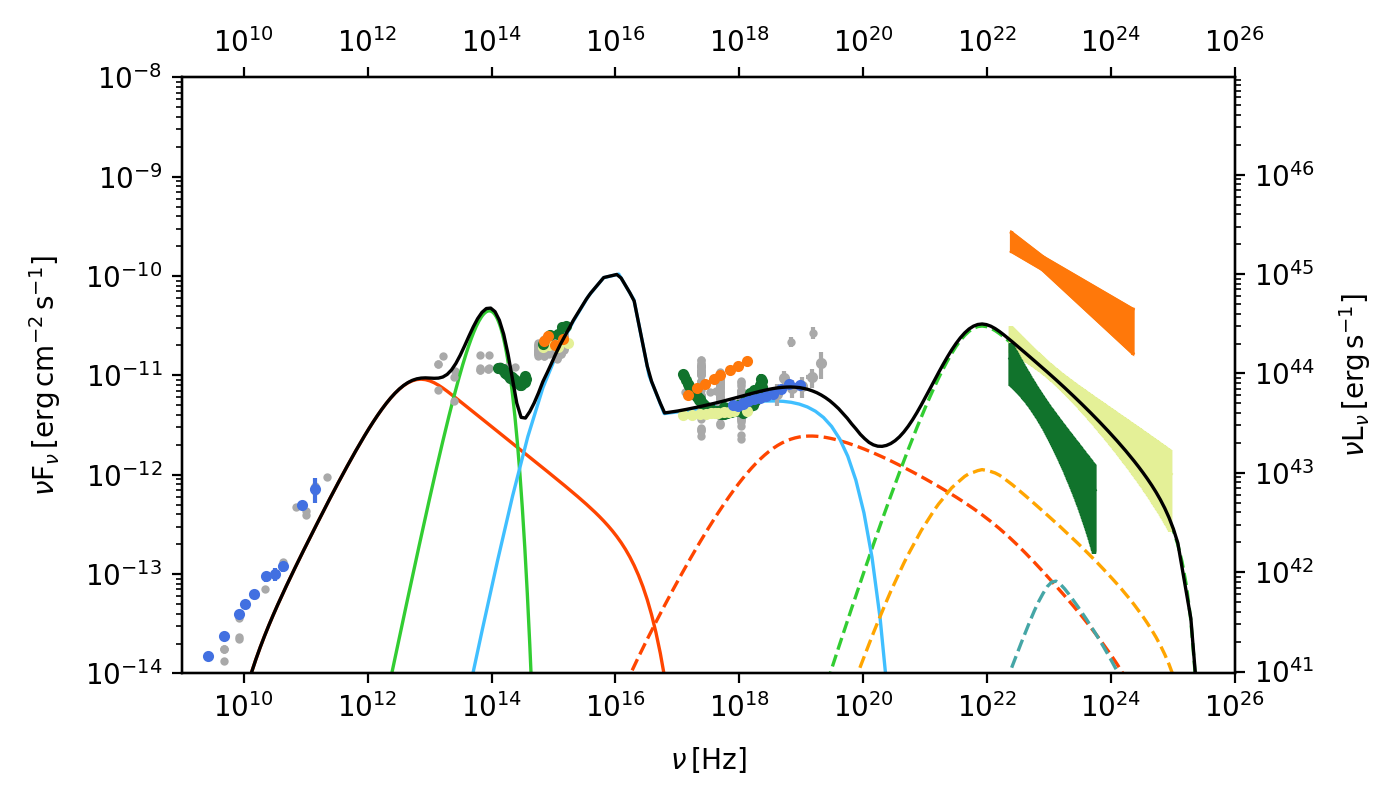}
     \end{subfigure}
     
     \begin{subfigure}
         \centering
         \includegraphics[width=0.7\textwidth]{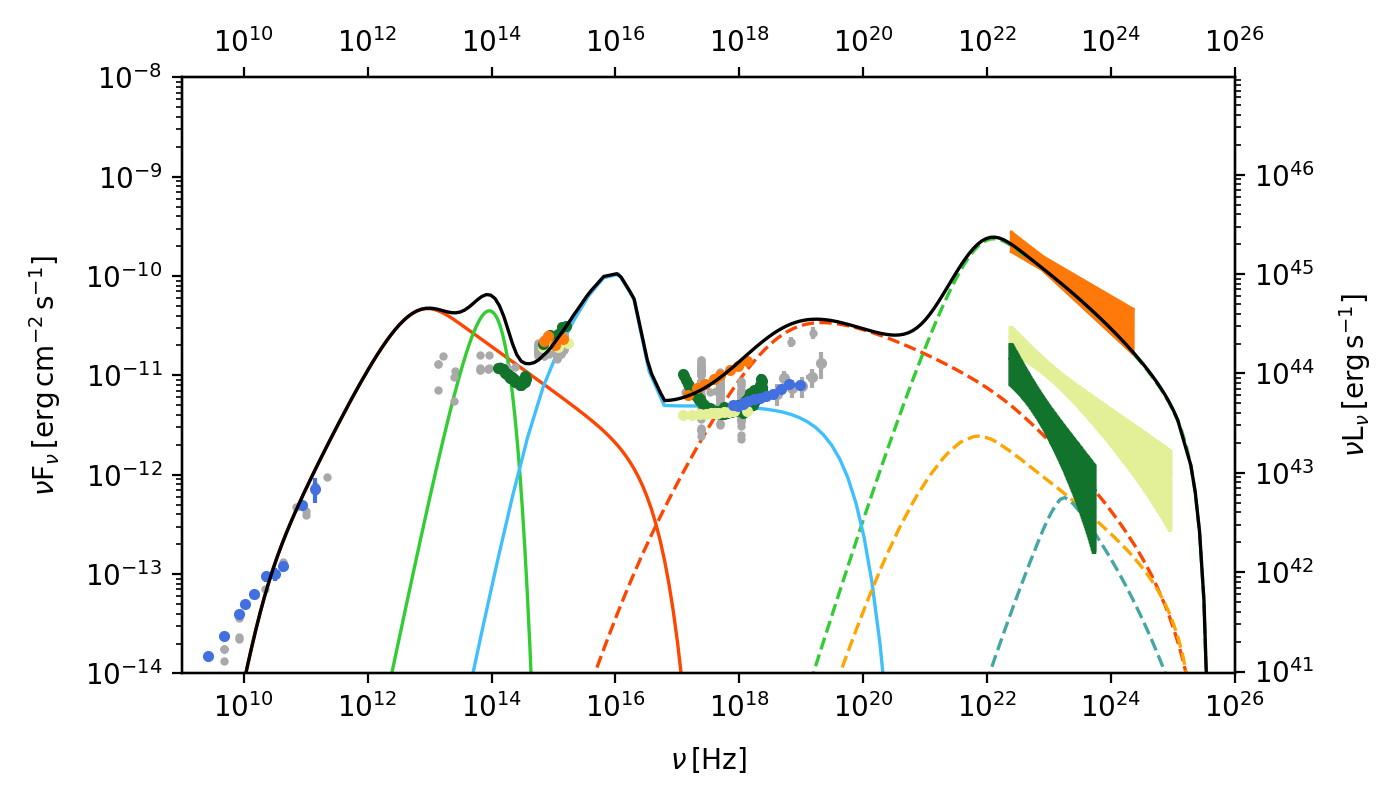}
     \end{subfigure}
        \caption{Torus dominated scenario for 1H 0323+342. The description of the model components and the data is the same as in Fig. \ref{fig:1H_Disc_BLR}.}
        \label{fig:1H_Torus}
\end{figure*}

\begin{figure*}
     \centering
     \begin{subfigure}
         \centering
         \includegraphics[width=0.7\linewidth]{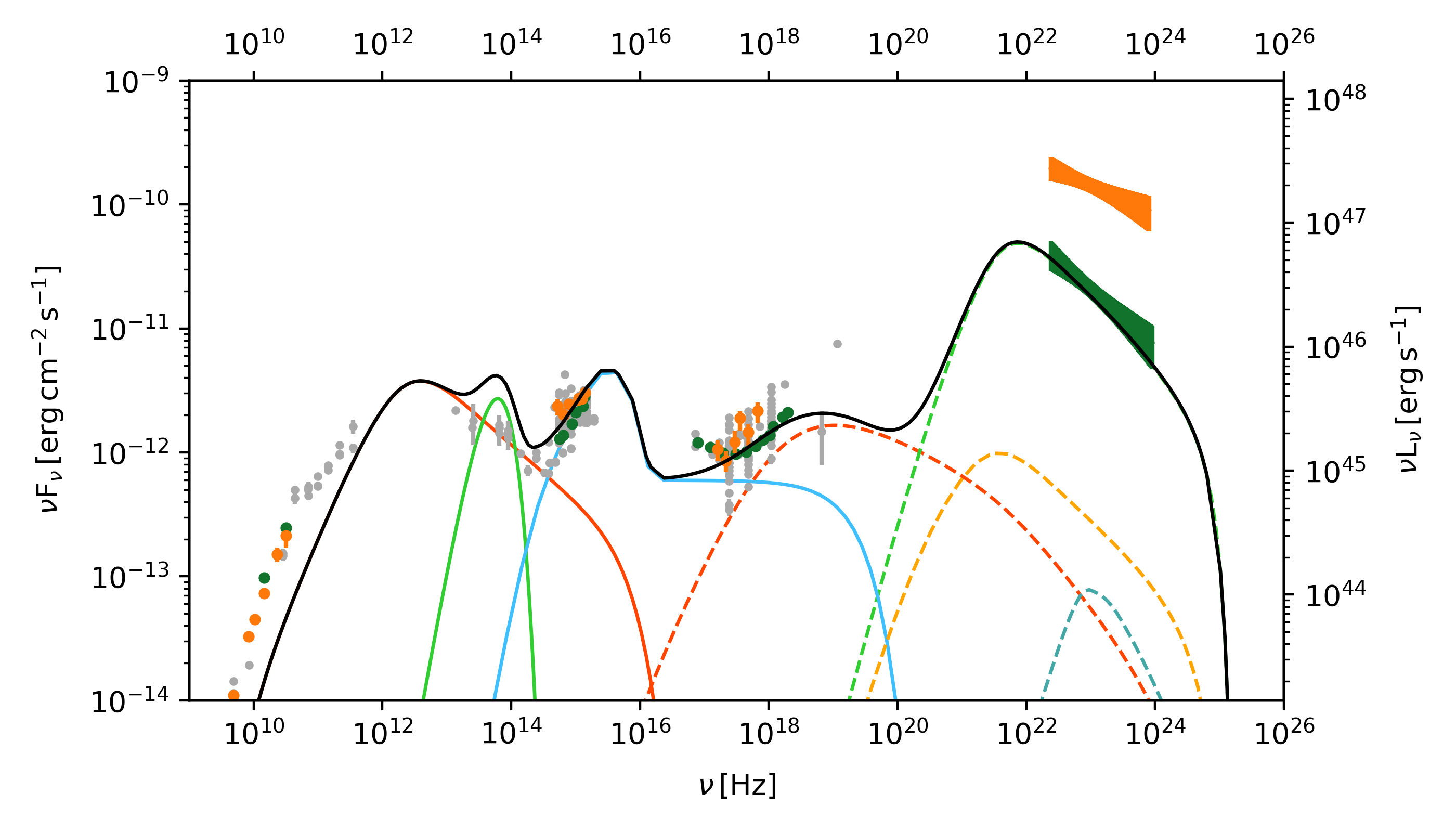}
     \end{subfigure}
     \begin{subfigure}
         \centering
         \includegraphics[width=0.7\textwidth]{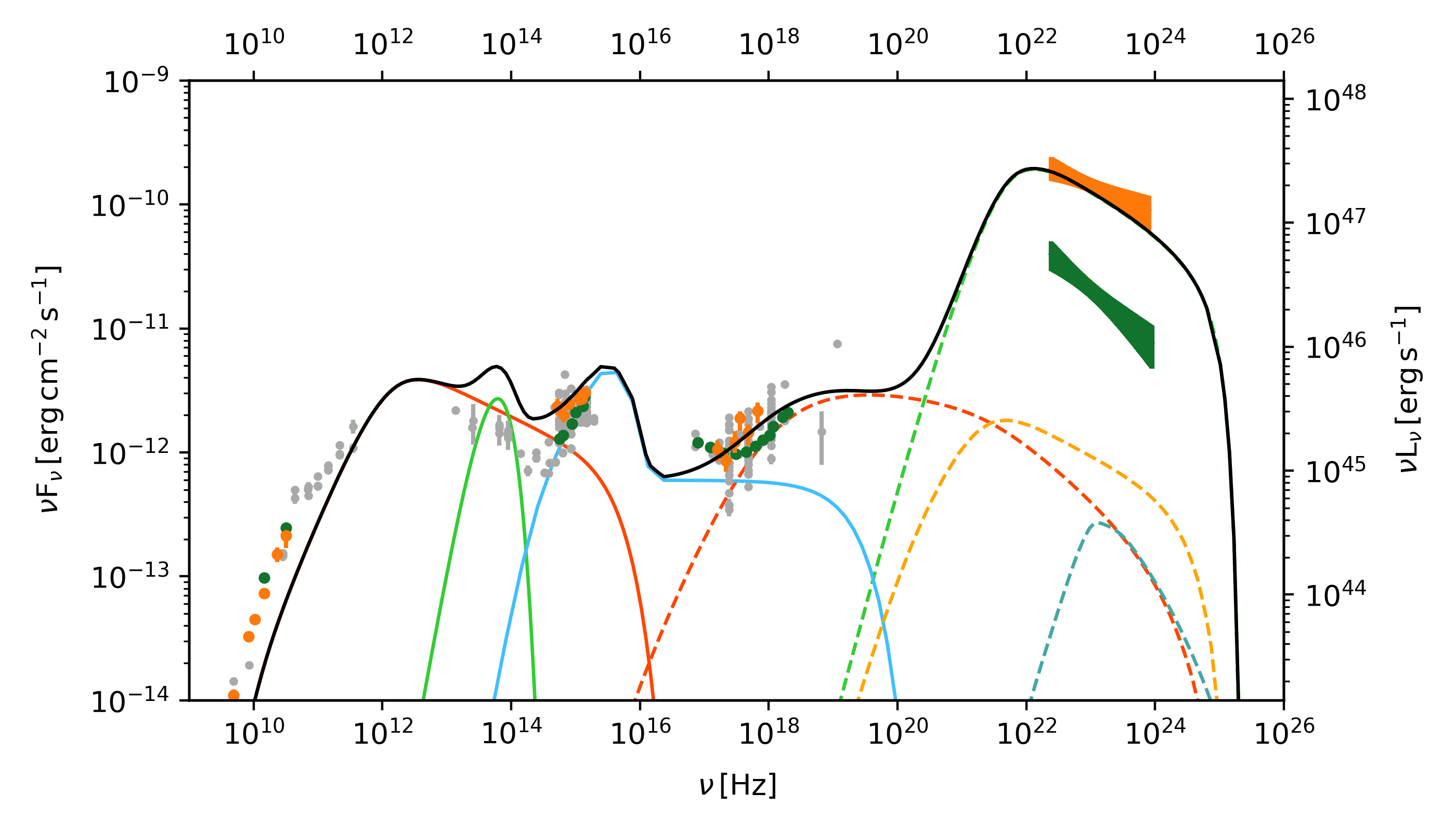}
     \end{subfigure}
        \caption{Torus dominated scenario for PMN J0948+0022. The description of the model components and the data is the same as in Fig. \ref{fig:1H_Disc_BLR}.}
        \label{fig:PMN_Torus}
\end{figure*}

\begin{figure*}
     \centering
     \begin{subfigure}
         \centering
         \includegraphics[width=0.7\linewidth]{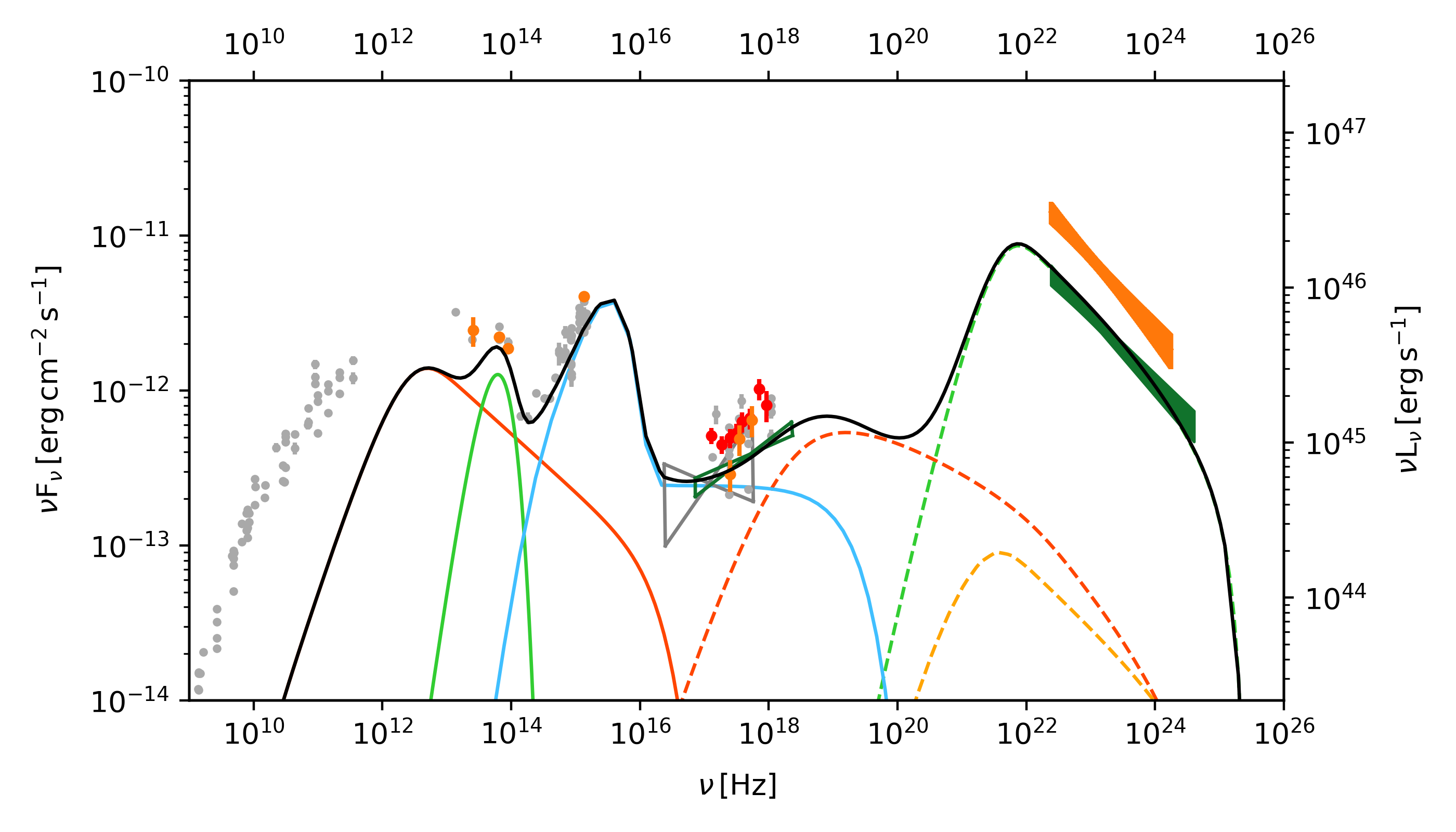}
     \end{subfigure}
     \begin{subfigure}
         \centering
         \includegraphics[width=0.7\textwidth]{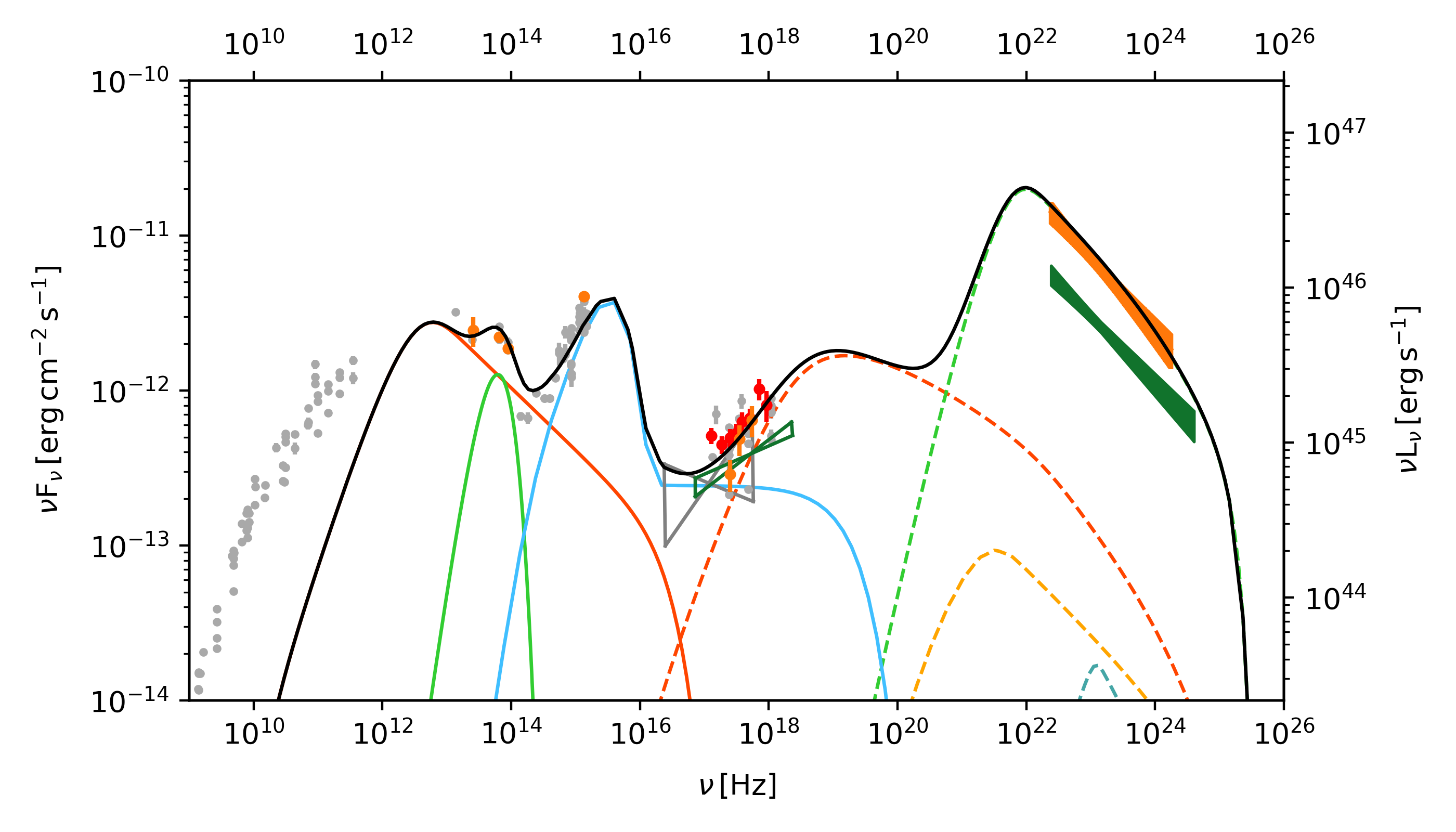}
     \end{subfigure}
        \caption{Torus dominated scenario for B2 0954+25A. The description of the model components and the data is the same as in Fig. \ref{fig:1H_Disc_BLR}.}
        \label{fig:B2_Torus}
\end{figure*}

\begin{figure*}
\centering
  $\vcenter{\hbox{\includegraphics[height=6cm]{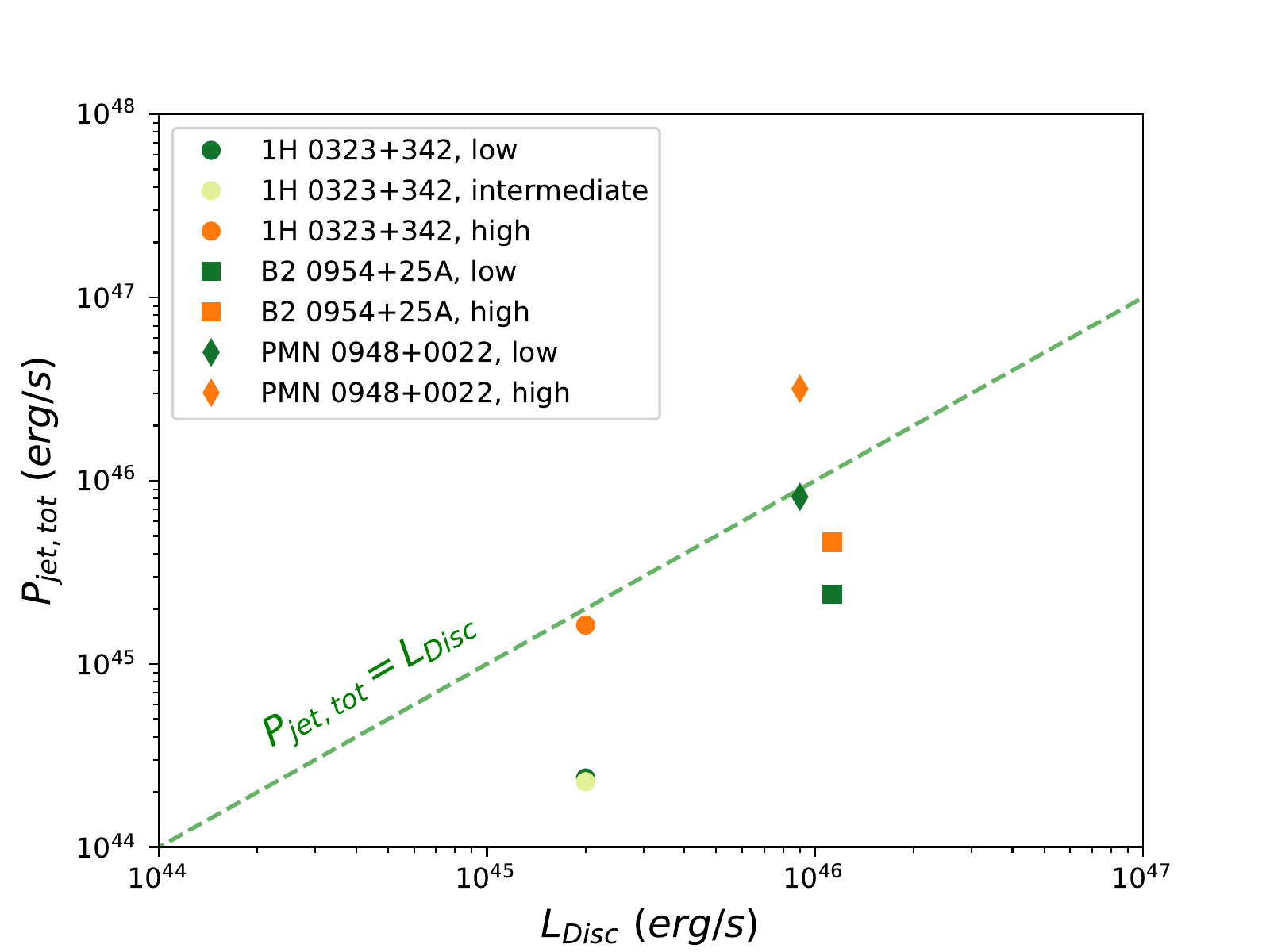}}}$
  \qquad
  $\vcenter{\hbox{\includegraphics[height=6cm]{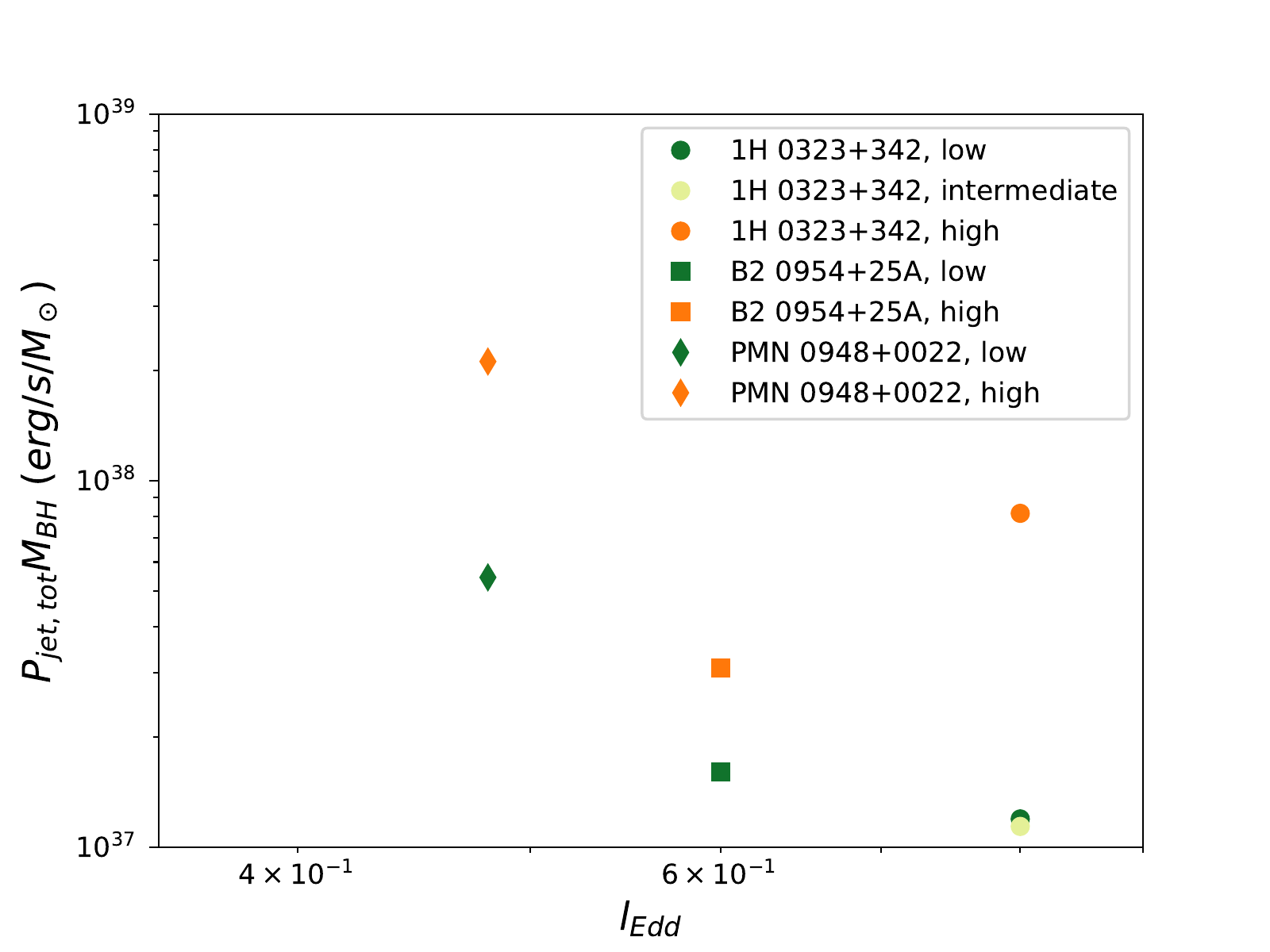}}}$
\caption{Total jet power vs disc luminosity (left panel) and total jet power corrected for BH mass vs Eddington ratio (right panel) for torus-dominated scenario.}
\label{fig:characterisation_upper_panel_torus}
\end{figure*}

\onecolumn
\section{Components of the radiative power for all models}

\FloatBarrier

\begin{table*}
    \centering
    \begin{tabular}{l|c|c|c|c|c|c|c|c}
          Source & Scenario & State & Syn &  SSC & EIC-Disc & EIC-Torus &  EIC-BLR & $P_{rad,tot}$ \\
                     \hline
                     \hline
         \multirow{6}{*}{1H 0323+342} & \multirow{3}{*}{Disc-BLR} & Low  & $3.05\times10^{41}$ & $8.04\times10^{40}$ & $7.19\times10^{42}$ & $6.95\times10^{40}$ & $5.57\times10^{43}$ & $6.33\times10^{43}$ \\
                                      &                           & Intermediate & $5.05\times10^{41}$ & $2.08\times10^{41}$ & $1.11\times10^{43}$ & $1.08\times10^{41}$ & $8.06\times10^{43}$ & $9.26\times10^{43}$ \\
                                      &                           & High & $2.73\times10^{42}$ & $4.13\times10^{42}$ & $3.66\times10^{43}$ & $6.99\times10^{41}$ & $4.05\times10^{44}$ & $4.50\times10^{44}$ \\
                                      & \multirow{3}{*}{Torus}    & Low  & $1.49\times10^{42}$ & $5.38\times10^{41}$ & $3.23\times10^{41}$ & $6.80\times10^{43}$ & $1.23\times10^{41}$ & $7.04\times10^{43}$ \\
                                      &                           & Intermediate & $1.77\times10^{42}$ & $7.31\times10^{41}$ & $3.73\times10^{41}$ & $7.84\times10^{43}$ & $1.25\times10^{41}$ & $8.14\times10^{43}$ \\
                                      &                           & High & $1.01\times10^{43}$ & $1.14\times10^{43}$ & $9.08\times10^{41}$ & $7.94\times10^{44}$ & $1.11\times10^{42}$ & $8.18\times10^{44}$ \\
         \hline

         \multirow{4}{*}{PMN J0948+0022} & \multirow{2}{*}{Disc-BLR} & Intermediate  & $5.41\times10^{42}$ & $1.99\times10^{42}$ & $6.87\times10^{43}$ & $1.33\times10^{43}$ & $1.08\times10^{46}$ & $1.09\times10^{46}$ \\
                                      &                              & High & $3.02\times10^{43}$ & $3.99\times10^{43}$ & $5.19\times10^{44}$ & $7.56\times10^{43}$ & $4.10\times10^{46}$ & $4.17\times10^{46}$ \\
                                      & \multirow{2}{*}{Torus}       & Intermediate  & $5.20\times10^{43}$ & $3.47\times10^{43}$ & $2.37\times10^{43}$ & $6.55\times10^{45}$ & $6.34\times10^{42}$ & $6.67\times10^{45}$ \\
                                      &                              & High & $5.05\times10^{43}$ & $5.85\times10^{43}$ & $4.04\times10^{43}$ & $2.85\times10^{46}$ & $2.18\times10^{43}$ & $2.87\times10^{46}$ \\
        \hline
        
         \multirow{4}{*}{B2 0954+25A} & \multirow{2}{*}{Disc-BLR} & Low  & $1.61\times10^{43}$ & $5.66\times10^{42}$ & $3.87\times10^{43}$ & $8.32\times10^{42}$ & $1.32\times10^{45}$ & $1.39\times10^{45}$ \\
                                      &                           & High & $2.26\times10^{43}$ & $7.95\times10^{42}$ & $5.69\times10^{43}$ & $1.95\times10^{43}$ & $3.22\times10^{45}$ & $3.33\times10^{45}$ \\
                                      & \multirow{2}{*}{Torus}    & Low  & $3.05\times10^{43}$ & $1.90\times10^{43}$ & $3.39\times10^{42}$ & $1.64\times10^{45}$ & $8.00\times10^{41}$ & $1.70\times10^{45}$ \\
                                      &                           & High & $4.35\times10^{43}$ & $4.27\times10^{43}$ & $2.55\times10^{42}$ & $3.26\times10^{45}$ & $1.50\times10^{43}$ & $3.35\times10^{45}$ \\

        \hline
    \end{tabular}
    \caption{Different contributions to the total radiative power from synchrotron, SSC, EIC of the disc, torus and BLR photons. }
    \label{tab:rad_contributions}
\end{table*}

\bsp	
\label{lastpage}
\end{document}